\documentclass[aps,superscriptaddress,showpacs,pre,10pt,nofootinbib,twocolumn]{revtex4-2}
\usepackage{bm} 
\usepackage{graphicx} 
\usepackage{amsmath}
\usepackage{amsthm}
\usepackage{amssymb} 
\usepackage{epstopdf} 
\usepackage{comment}
\usepackage{amsfonts}
\usepackage{epsfig}
\usepackage{color} 
\usepackage[dvipsnames]{xcolor}
\usepackage{subfigure}
\usepackage[english]{babel}
\usepackage[bb=boondox]{mathalfa}
\usepackage{mathtools}
\usepackage{dsfont}
\usepackage{hyperref}
\usepackage[normalem]{ulem} 
\hypersetup{
    colorlinks=true,       
    linkcolor=cyan,          
    citecolor=magenta,        
    filecolor=magenta,      
    urlcolor=cyan,           
    runcolor=cyan
}

\newcommand{\ket}[1]{| #1 \rangle}

\newcommand {\be}{\begin{equation}}
\newcommand {\ee}{\end{equation}}

\newcommand{\ba}{\begin{eqnarray}}
\newcommand{\ea}{\end{eqnarray}}
\newcommand\tr{{\mbox{Tr\,}}}
\newcommand{\ignore}[1]{}

\usepackage[margin=1in,nofoot]{geometry}

\newcommand{\e}{{{e}}}

\renewcommand{\Re}{\operatorname{Re}}

\newcommand{\beq}{\begin{equation}}
\newcommand{\eeq}{\end{equation}}
\newcommand{\beqnn}{\begin{equation*}}
\newcommand{\eeqnn}{\end{equation*}}
\newcommand{\bea}{\begin{eqnarray}}
\newcommand{\eea}{\end{eqnarray}}
\newcommand{\beann}{\begin{eqnarray*}}
\newcommand{\eeann}{\end{eqnarray*}}
\newcommand{\bes} {\begin{subequations}}
\newcommand{\ees} {\end{subequations}}

\newcommand{\sgn}{\textrm{sgn}}
\newenvironment{claim}[1]{\par\noindent{\bf{Claim}:}\space#1}{}
\newenvironment{claimproof}[1]{\par\noindent{\bf{Proof}:}\space#1}{\hfill $\blacksquare$}

\begin{document}
\raggedbottom 
\title{A quantum Monte Carlo algorithm for arbitrary \texorpdfstring{spin-$1/2$}{spin-1/2} Hamiltonians}
\author{Lev Barash}
\email{lbarash@usc.edu}
\affiliation{Information Sciences Institute, University of Southern California, Marina del Rey, California 90292, USA}
\author{Arman Babakhani}
\affiliation{Information Sciences Institute, University of Southern California, Marina del Rey, California 90292, USA}
\affiliation{Department of Physics and Astronomy, University of Southern California, Los Angeles, California 90089, USA}
\author{Itay Hen}
\affiliation{Information Sciences Institute, University of Southern California, Marina del Rey, California 90292, USA}
\affiliation{Department of Physics and Astronomy, University of Southern California, Los Angeles, California 90089, USA}

\date{\today}

\begin{abstract}
\noindent 
We present a universal parameter-free quantum Monte Carlo (QMC) algorithm designed to simulate arbitrary spin-$1/2$ Hamiltonians.
To ensure the convergence of the Markov chain to equilibrium for every conceivable case, we devise a clear and simple automated protocol that produces QMC updates
that are provably ergodic and satisfy detailed balance.
We demonstrate the applicability and versatility of our method by considering several illustrative examples,
including the simulation of the XY model on a triangular lattice, the toric code, and random $k$-local Hamiltonians.
We have made our program code freely accessible on GitHub.
\end{abstract}

\maketitle

\section{Introduction}
Quantum Monte Carlo (QMC) algorithms~\cite{bookGubernatisQMC,bookLandauBinder,reviewLuchow2011} are an indispensable tool in the study of the equilibrium properties of large quantum many-body systems, with applications ranging from superconductivity and novel quantum materials~\cite{Qchem1,Qchem2,Qchem3}  through the physics of neutron stars~\cite{1742-6596-529-1-012012} and quantum chromodynamics~\cite{PhysRevLett.83.3116,PhysRevE.49.3855}.  The algorithmic development of QMC remains an active area of research, with continual efforts being made to extend the scope of QMC applicability and to improve convergence rates of existing algorithms with the goal of expanding our understanding of quantum systems and facilitating the discovery of novel quantum phenomena~\cite{sandvik:92,sandvik:99,prokofiev:98,Troyer2015,Li2015}. 

While QMC algorithms have been adapted to the simulation of a wide variety of physical systems, different physical models typically require the development of distinct model-specific update rules and measurement schemes, making the simulation of many large-scale quantum many-body systems prohibitively time-consuming.

In this paper, we introduce a universal parameter-free QMC algorithm designed to reliably simulate \emph{arbitrary} spin-$1/2$ Hamiltonians. To achieve such capabilities, we devise a clear and simple automated protocol for generating the necessary set of updates to ensure the ergodic Markov chain Monte Carlo sampling of any conceivable input system. The generated QMC updates are shown to be ergodic as well as satisfying detailed balance thereby guaranteeing the proper convergence of the QMC Markov chain. We demonstrate the validity and flexibility of our code by studying in detail a number of models, including the XY model on a triangular lattice, the toric code, random $k$-local Hamiltonians and more.
We have also made our program code freely accessible on GitHub~\cite{githubPMRQMC}.

The technique we propose here builds on the recently introduced Permutation Matrix Representation (PMR) QMC~\cite{pmr} -- an abstract Trotter-error-free technique wherein the quantum dimension consists of products of elements of permutation groups and which as a result allows for the general treatment of entire classes of Hamiltonians. In PMR QMC, the partition function is expanded in a power series in the off-diagonal strength of the Hamiltonian about the partition function of the classical (diagonal) component of the Hamiltonian~\cite{ODE,ODE2}.  

The paper is organized as follows. In Sec.~\ref{sec:perMatRep}, we provide a brief overview of the off-diagonal series expansion for quantum partition functions, on which our approach is founded. In Sec.~\ref{sec:Hamiltonians}, we analyze spin-$1/2$ Hamiltonians in the context of PMR-QMC. In the following section, Sec.~\ref{sec:qmc}, we discuss our method to generate all the necessary QMC updates, proving that these ensure both ergodicity and detailed balance. Section~\ref{sec:measurement} is devoted to the implementation of comprehensive measurement schemes. 
In Sec.~\ref{sec:Results} we showcase the power of our technique by detailing the simulation results of the physical models mentioned above. We conclude in Sec.~\ref{sec:conclusions} with an additional discussion of future directions of research.

\section{The off-diagonal partition function expansion}\label{sec:perMatRep}

The permutation matrix representation (PMR) protocol~\cite{pmr} begins by first casting the to-be-simulated Hamiltonian in PMR form, i.e., as a sum
\beq \label{eq:basic}
H=\sum_{j=0}^M \tilde{P}_{j} =\sum_{j=0}^M D_j P_j =D_0 + \sum_{j=1}^M D_j P_j \,,
\eeq
where $\{ \tilde{P}_j\}$ is a set of $M+1$ distinct generalized permutation matrices~\cite{gpm}, i.e., matrices with at most one nonzero element in each row and each column. Each operator $\tilde{P}_j$ can be written, without loss of generality, as $\tilde{P}_j=D_j P_j$ where $D_j$ is a diagonal matrix and $P_j$ is a  permutation matrix with no fixed points (equivalently, no nonzero diagonal elements) except for the identity matrix $P_0=\mathbb{1}$. We will refer to the basis in which the operators $\{D_j\}$ are diagonal as the computational basis and denote its states by $\{ |z\rangle \}$. We will call the diagonal matrix $D_0$ the `classical Hamiltonian'.
The permutation matrices appearing in $H$ will be treated as a subset of a permutation group, wherein $P_0$ is the identity element. 

The $\{D_j P_j \}$  off-diagonal operators (in the computational basis) give the system its  `quantum dimension'.  For $j>0$, each term $D_j P_j$ obeys 
$D_j P_j | z \rangle = d_j(z') | z' \rangle$ where $d_j(z')$ is a possibly complex-valued coefficient and $|z'\rangle \neq |z\rangle$ is a basis state.

Upon casting the Hamiltonian in PMR form, one can show~\cite{pmr} that the partition function \hbox{$Z=\tr\left[ \e^{-\beta H} \right]$} can be written as
\beq\label{eq:SSE3}
Z  = \sum_{z} \sum_{q=0}^{\infty} \sum_{S_{{\bf{i}}_q}} D_{(z,S_{{\bf{i}}_q})} e^{-\beta[E_{z_0},\ldots,E_{z_q}]} \langle z | S_{{\bf{i}}_q} | z \rangle \,.
\eeq
The sum above is a double sum: over the set of all basis states $z$ and over all products of $q$ off-diagonal permutation operators  $S_{{\bf{i}}_q}=P_{i_q} \ldots P_{i_2}P_{i_1}$ with $q$ running from zero to infinity. The multi-index \hbox{${\bf{i}}_q=(i_1,\ldots,i_q)$} covers all products of permutation operators, where each index $i_j$ runs from $1$ to $M$.

In the above sum, each summand is a product of three terms. The first is $D_{ ( z, S_{\mathbf{i}_q} ) } \equiv \prod_{j=1}^{q} d_{z_j}^{(i_j)}$ consisting of a product of the matrix elements
\begin{equation}
d_{z_j}^{(i_j)} = \langle z_j \vert D_{i_j} \vert z_j \rangle \,. 
\end{equation}
The various $\{|z_j\rangle\}_{j=0}^q$ states are the states obtained from the action of the ordered $P_{i_j}$ operators in the product $S_{{\bf{i}}_q}$ on $|z_0\rangle$, then on $|z_1\rangle$, and so forth. For $S_{{\bf{i}}_q}=P_{i_q} \ldots P_{i_2}P_{i_1}$, we have $|z_0\rangle=|z\rangle, P_{i_j}|z_{j-1}\rangle=|z_j\rangle$ for $j = 1,2,\dots,q$. The proper indexing of the states $|z_j\rangle$ is \hbox{$|z_{(i_1,i_2,\ldots,i_j)}\rangle$} to indicate that the state at the $j$-th step depends on all $P_{i_1}\ldots P_{i_j}$. We will however use the shorthand $|z_j\rangle$ to simplify the notation. The sequence of basis states $\{|z_j\rangle \}$ may be viewed as a `walk'~\cite{graphTheory} on the Hamiltonian graph where every matrix element $H_{ij}$ corresponds to an edge between the two basis states, or nodes, $i$ and $j$. 

The second term in each summand, $e^{-\beta[ E_{z_0}, \ldots, E_{z_q}]}$,  is called the divided differences of the function \hbox{$f(\cdot) = e^{-\beta (\cdot)}$} with respect to the inputs $[ E_{z_0}, \ldots, E_{z_q}]$, where $E_{z_i} = \langle z_i \vert H \vert z_i \rangle = \langle z_i \vert D_0 \vert z_i \rangle$. The divided differences~\cite{dd:67,deboor:05} of the function $f$ is defined as,\footnote{The expression, Eq.~(\ref{eq:divdiff}), is ill-defined if two (or more) of the inputs are repeated, in which case $f[E_{z_0},\ldots,E_{z_q}]$ can be properly evaluated using limits.}
\begin{equation}
f[ E_{z_0}, \ldots, E_{z_q} ] \equiv \sum_{j=0}^{q} \frac{ f(E_{z_j}) }{ \prod_{k \neq j} ( E_{z_j} - E_{z_k} ) }.
\label{eq:divdiff}
\end{equation}
A useful property is that the divided differences are invariant under rearrangements of the input values.

Now, the term $\langle z| S_{{\bf{i}}_q} |z\rangle$ in Eq.~(\ref{eq:SSE3}) evaluates either to 1 or to zero.
Moreover, since the permutation matrices with the exception of $P_0$ have no fixed points, the condition $\langle z| S_{{\bf{i}}_q} |z\rangle=1$ implies $S_{{\bf{i}}_q}=\mathbb{1}$, i.e., $S_{{\bf{i}}_q}$ must evaluate to the identity element $P_0$ (note that the identity element does not appear in the sequences $S_{{\bf{i}}_q}$).  The expansion Eq.~(\ref{eq:SSE3}) can thus be more succinctly rewritten as
 \beq\label{eq:z}
Z  =\sum_{z}\sum_{S_{{\bf{i}}_q}=\mathbb{1}}  D_{(z,S_{{\bf{i}}_q})}   \e^{-\beta [E_{z_0},\ldots,E_{z_q}]} \,,
\eeq
i.e., as a sum over all basis states and permutation matrix products that evaluate to the identity matrix.

Having derived the expansion Eq.~(\ref{eq:z}) for any Hamiltonian cast in the form Eq.~(\ref{eq:basic}), we are now in a position to interpret the partition function expansion as a sum of weights, i.e., $Z = \sum_{{{\cal C}}} w_{{\cal C}}$, where the set of configurations $\{{\cal C}\}$ consists of all the distinct pairs $\{ |z\rangle, S_{{\bf{i}}_q} \}$. 
We refer to
\beq
w_{{\cal C}} =  D_{(z,S_{{\bf{i}}_q})}  \e^{-\beta [E_{z_0},\ldots,E_{z_q}]}
\eeq
as the configuration weight.  
We note that as written, the weights $w_{\cal C}$ can in general be complex-valued, despite the partition function $Z = \sum_{\cal C} w_{\cal C}$ being real (and positive).
Hence, the imaginary portions of the complex-valued weights do not contribute to the partition function
and may be disregarded altogether.\footnote{In fact, it can be shown that for every configuration \hbox{$\mathcal{C}=\{ |z\rangle, P_{i_q} \ldots  P_{i_2} P_{i_1}\}$} the configuration with reverse order of operators $\overline{\mathcal{C}}=\{ |z\rangle, P_{i_1}^{-1} P_{i_2}^{-1}\ldots  P_{i_q}^{-1}\}$ produces the conjugate weight $w_{\overline{\mathcal{C}}}=\overline{w}_{\mathcal{C}}$.}
We may therefore ignore the imaginary components of the weights to obtain the strictly real-valued weights
\beq
\label{eq:gbw}
W_{{\cal C}}=\Re\left[D_{(z,S_{{\bf{i}}_q})}\right]  \e^{-\beta [E_{z_0},\ldots,E_{z_q}]}\,.
\eeq

\section{Permutation matrix representation of \texorpdfstring{Spin-$1/2$}{Spin-1/2} Hamiltonians}
\label{sec:Hamiltonians}

We next discuss the PMR formulation of general spin-$1/2$ Hamiltonians, which are the main focus of this work. 
Hamiltonians of spin-1/2 systems are often, and can always be, represented using Pauli matrices ($X$, $Y$, and $Z$). In order to represent this class of Hamiltonians in PMR form, Eq.~(\ref{eq:basic}), we choose our group of permutations to be the set of all tensor products of Pauli-$X$ operators, $G_X$, while the diagonal matrices $D_j$ will be expressed in terms of tensor products of Pauli-$Z$ matrices. As we shall see, any permutation operator for spin-$1/2$ systems belongs to $G_X$.

Consider a Hamiltonian $H$ given as a linear combination of Pauli strings  $S^{(i)} = \bigotimes_j s^{(i)}_j$.
Here, $s^{(i)}_j$ represents a Pauli matrix $s \in \{X,Y,Z\}$ in the $i$-th Pauli string, which acts on the $j$-th spin, where $j \in \{1, \ldots, n\}$ with $n$ being the number of spins in the system (we assume that each Pauli string $S^{(i)}$ contains no more than one Pauli matrix $s_j^{(i)}$ for each spin index $j$). Explicitly, the Hamiltonian has the form
\beq
H=\sum_{i} c_i S^{(i)} \,,
\label{eq:PauliStrings}
\eeq
where the $c_i$'s are real-valued coefficients. To cast the Hamiltonian in PMR form, we first write each Pauli string as a product of a diagonal operator and a permutation operator, i.e. we will write $S^{(i)}$ as a string of Pauli-$Z$'s multiplied by a string of Pauli-$X$'s using the fact that $Y = -i Z X$ giving
\beq
H =\sum_i \tilde{c}_i  Z^{(i)}  X^{(i)} \,,
\eeq
where $\tilde{c}_i = (-i)^{n_y^{(i)}}c_i$ with  $n_y^{(i)}$ being the number of $Y$ operators appearing in the $i$-th Pauli string. Here, $X^{(i)}$ represents the $i$-th string (or product) of Pauli-$X$ operators. The notation  $Z^{(i)}$ similarly represents a string of Pauli-$Z$ operators. 

Noting that Pauli-$X$ strings are permutation operators and that Pauli-$Z$ strings are diagonal in our chosen basis, we next group together all terms that have the same Pauli-$X$ component, ending up with a Hamiltonian of the form 
\beq
H= \sum_i \left( \sum_j \tilde{c}_j Z^{(i)}_j \right) X^{(i)}.
\eeq
The permutation operators $X^{(i)} \in G_X$ in the above expression are now \emph{distinct} products of Pauli-$X$ operators, and we identify \hbox{$D_i=\sum_j \tilde{c}_j Z^{(i)}_j$} as the accompanying diagonal operators, as desired.

\section{QMC updates for arbitrary \texorpdfstring{spin-$1/2$}{spin-1/2} Hamiltonians}\label{sec:qmc}

\subsection{QMC configurations} 

As was discussed above, for any Hamiltonian cast in PMR form, the partition function $Z=\tr[e^{-\beta H}]$ can be written as a sum of configuration weights, where a configuration \hbox{${\cal C}=\{|z\rangle, S_{{\bf{i}}_q}\}$} is a pair of 
a `classical' basis state $|z\rangle$ (an eigenstate of diagonal operators) and a product $S_{{\bf{i}}_q}$ of permutation operators that must evaluate to the identity element $P_0=\mathbb{1}$. 
The configuration ${\cal C}$ induces a list of states \hbox{$\{|z_0\rangle =|z\rangle ,|z_1\rangle ,\ldots,|z_q\rangle =|z\rangle \}$}, which in turn generates a corresponding multi-set of
energies $E_{{\cal C}}=\{E_{z_0},E_{z_1},\ldots,E_{z_q}\}$ for the configuration.

We can now consider a QMC algorithm, i.e., a Markov Chain Monte Carlo routine, that samples these configurations with probabilities proportional to their weights $W_{{\cal C}}$, Eq.~(\ref{eq:gbw}). The Markov process would start with some initial configuration and a set of (probabilistic) rules, or QMC updates, will dictate transitions from one configuration to the next. 

We will take the initial state of the chain to be 
\hbox{$\mathcal{C}_0=\{|z\rangle, S_0=\mathbb{1}\}$} where $|z\rangle$ is a randomly generated initial classical state.
The weight of this initial configuration is 
\beq
W_{{\cal C}_0}=\e^{-\beta [E_z]}=\e^{-\beta E_z} \,,
\eeq
i.e., the classical Boltzmann weight of the initial randomly generated basis state $|z\rangle$. 

The set of QMC updates will be discussed in the next sections. Before doing so, we note that to make certain that configurations are sampled properly, i.e. in proportion to their weight, we will ensure that the Markov chain is (i) ergodic, i.e., that the QMC updates are capable of generating all basis states $|z\rangle$ as well as all sequences $S_{{\bf{i}}_q}$ evaluating to the identity, and (ii) satisfies detailed balance, a sufficient (but not strictly necessary) condition dictating that the ratio of transition probabilities from one configuration to another and the transition in the opposite direction equals to the ratio of their respective weights~\cite{bookNewmanBarkema,bookGrimmett}. 
In what follows, we show how both conditions are made to be satisfied. 

\subsection{Fundamental cycles}
\label{sec:fundamentalcycles}

As noted above, for spin-$1/2$ Hamiltonians, the permutation operators $P_i$ are Pauli-$X$ strings. As such, they obey (i) $[P_i,P_j]=0$ for every $i,j$, i.e., they all commute and (ii) $P_i^2=\mathbb{1}$. Moreover, each Pauli-$X$ string can be represented as a bit-string of the form $[b_1b_2 \cdots b_n]$ where the $i$-th bit $b_i \in \{0,1\}$ indicates whether $X_i$ is in the string ($b_i=1$) or not ($b_i=0$). (For example, $[1 \, 0 \, 1 \, 0 \, 0 \, 1 ]$ would correspond to the string $X_1 X_3 X_6$.) Denoting by $p_i$ the bit-string corresponding to the operator $P_i$, one can easily verify that the product of two permutation operators $P_i$ and $P_j$ would likewise correspond to the addition modulo 2 (the XOR operation) of the two respective bit-strings $p_i \oplus p_j$. Specifically, a sequence of operators evaluating to the identity, $S_{{\bf{i}}_q}=P_{i_q} \ldots P_{i_2}P_{i_1}=\mathbb{1}$, can be written as $\oplus_{j=1}^q p_{i_j} = {\bf 0}$, where ${\bf 0}$ is the zero bit-string -- the bit-string consisting of only zeros.

Moreover, the following observations may be made: (i) Any given sequence of permutation operators $S_{{\bf{i}}_q}$ that evaluates to the identity is a permutation of a multi-set of operators, the product of which evaluates to the identity. (ii) A multi-set of operators obtained from another by the removal of a pair of identical operators will evaluate to the identity if and only if the original multi-set evaluates to the identity (this follows from the commutativity of permutation operators and the fact that $P_i^2=\mathbb{1}$). (iii) Upon removing all pairs of identical operators from a multi-set, one is left with a (proper) set of permutation operators, the product of which evaluates to the identity. This set consists of the permutation operators that are contained an odd number of times in the original sequence $S_{{\bf{i}}_q}$.

We shall call a set of permutation operators that multiply to the identity a `cycle'. In terms of binary strings, cycles may be represented by bit-strings $[a_1a_2 \cdots a_M]$ indicating which of the permutation operators $P_1, \dots, P_M$ belong to the set. The length of a cycle would be the number of permutation operators in it (equivalently, the number of $1$'s in its binary vector representation). As pointed out above, any sequence of operators $S_{{\bf{i}}_q}$ has a `core' cycle from which $S_{{\bf{i}}_q}$ can be generated via insertions of operator pairs followed by reordering. 

The question of how one can generate all possible sequences of operators (which evaluate to the identity) is thus reduced to the question of how one can generate all possible cycles (as the former can be derived from the latter).
In terms of binary strings, the above question translates to generating the bit-strings $[a_1a_2 \cdots a_M]$ ($a_j \in \{0,1\}$) that obey the following system of linear equations over mod-$2$ addition: $\oplus_{j=1}^M a_j p_j = {\bf 0}$.
This question can be answered by finding the nullspace over mod-$2$ addition of the matrix whose columns are the bit-strings $p_j$.
Any bit-string $[a_1a_2 \cdots a_M]$ obeying the condition $\oplus_{j=1}^M a_j p_j = {\bf 0}$ can be written as mod-$2$ addition of bit-strings from the nullspace basis. 

Now, the nullspace basis can be found efficiently via Gaussian elimination over mod-2 addition~\cite{GaussianElimination1991}.
We shall refer to cycles represented by the bit-strings from the nullspace basis as fundamental cycles. Any cycle is therefore a combination of fundamental cycles.

Generally, the nullspace basis states can be chosen in many different ways, and so the choice of the set of fundamental cycles is not unique.
From a practical point of view, however, we find that obtaining a `minimal cycle basis', i.e., a basis that minimizes the lengths of all basis cycles, is advantageous from the QMC standpoint. This follows from the fact that the probability of a QMC update to be accepted is a decreasing function of the cycle length (see Sec.~\ref{subsec:QMCupdates}). 
To reduce the cycles lengths, we therefore find a basis using Gaussian elimination and then proceed to replace long-cycle basis states with shorter basis states by performing mod-$2$ additions between the bit-strings of pairs of cycles,
accepting the changes each time a new cycle with a shorter length is found. 
The process ends when a pass through all pairs of cycles does not result in an improvement.

It is worth noting that minimizing cycle lengths of Hamiltonians with nontrivial topologies will not always lead to a cycle basis whose fundamental cycle lengths are all of the order $O(1)$, i.e., do not grow with system size. In the presence of a non-trivial topology, e.g., where periodic boundary conditions are imposed or for other nonzero genus models, there may exist fundamental cycles becoming longer with the system size such as the cycles `wrapping around the system', consisting of an extensive number of permutation operators. 

\subsection{Ergodicity\label{sec:ergodicity}}

We are now in a position to utilize the insights from the previous subsection to establish a set of updates to the sequence of operators $S_{\boldsymbol{i}_q}$ that would allow for the possibility of generating all possible operator sequences that evaluate to the identity. 

As a preliminary step, prior to the simulation taking place, we produce a list of fundamental cycles $\mathcal{C}_1, \dots, \mathcal{C}_T$, where $T$ is the dimension of the nullspace, for the to-be-simulated Hamiltonian cast in PMR form. 

We next prove that all possible sequences can be generated, i.e., that the Markov chain is ergodic, via a combination of insertion and removal of fundamental cycles, changing the order of the permutation operators in the sequence (permutation operator swapping), and the insertion and removal of pairs of identical permutation operators.

\begin{claim}
    Every permutation operator sequence evaluating to the identity can be generated by
    (i) insertion and removal of fundamental cycles,
     (ii) insertion and removal of pairs of identical permutation operators,
     (iii) swapping the order of two adjacent permutation operators.
\end{claim}

\begin{claimproof}
For each permutation operator sequence $S_{\boldsymbol{i}_q}$ evaluating to the identity, one can consider
the cycle $\mathcal{C}_{\boldsymbol{i}_q}$ consisting of permutation operators 
that appear an odd number of times in $S_{\boldsymbol{i}_q}$. In other words, by removal of pairs of identical operators, the multi-set of operators that are contained in $S_{\boldsymbol{i}_q}$ can be reduced to the cycle $\mathcal{C}_{\boldsymbol{i}_q}$ consisting of permutation operators that appear only once.

As shown in Sec.~\ref{sec:fundamentalcycles}, the cycle $\mathcal{C}_{\boldsymbol{i}_q}$ 
can be expressed through fundamental cycles, since any cycle can be expressed in terms of the mod-2 nullspace of the matrix containing bit-strings of permutation operators $P_i$ from~(\ref{eq:basic}).
Specifically, $c_{\boldsymbol{i}_q} = \oplus_{j=1}^T \alpha_j \mu_j$, where
$c_{\boldsymbol{i}_q}$ is the bit-string representation of the cycle $\mathcal{C}_{\boldsymbol{i}_q}$,
$\mu_j$ is a bit-string representation of the fundamental cycle $\mathcal{C}_j$, and
$\alpha_j \in \{0,1\}$, $j = 1,2,\dots,T$.

The latter observation implies that the cycle $\mathcal{C}_{\boldsymbol{i}_q}$ can be obtained by
inserting those fundamental cycles $\mathcal{C}_j$ for which the condition $\alpha_j=1$ is satisfied,
accompanied by any necessary reordering of permutation operators and the removal of pairs of identical permutation operators.
In turn, the sequence $S_{\boldsymbol{i}_q}$ can be obtained from $\mathcal{C}_{\boldsymbol{i}_q}$
via the addition of pairs of permutation operators together with reordering of permutation operators. This concludes the proof of our claim.
    
\end{claimproof}

The ability to generate all permutation operator sequences that evaluate to the identity, $S_{\boldsymbol{i}_q}$, implies ergodicity along the quantum (or imaginary time) dimension. In addition, the generation of all possible classical basis states $|z\rangle$ can be achieved independently by considering moves such as single- or multi-spin flips of the basis states. Moves of this nature guarantee ergodicity along the `classical' dimension.  

In the next section we will use the above observations to devise QMC updates that  ensure  ergodicity in the entire configuration space, which is the direct product of the classical and quantum configuration spaces, i.e., the pairs ${\cal C} = \{ |z\rangle, S_{{\bf{i}}_q} \}$.

\subsection{QMC updates} 
\label{subsec:QMCupdates}

We next describe the basic update moves for our QMC algorithm.
To ensure both performance and high accuracy of evaluating the weight of any given configuration $C=\{ |z\rangle, S_{{\bf{i}}_q} \}$,  we employ a divided differences 
calculation by means of addition and removal of items from the input multi-set of classical energies~\cite{GuptaBarashHen2020}.
We recall that the weight of a configuration requires the calculation of the divided differences $e^{-\beta[ E_{z_0}, \ldots, E_{z_q}]}$ with inputs $[ E_{z_0}, \ldots, E_{z_q}]$, where $E_{z_i} = \langle z_i \vert H \vert z_i \rangle = \langle z_i \vert D_0 \vert z_i \rangle$. For $S_{{\bf{i}}_q}=P_{i_q} \ldots P_{i_2}P_{i_1}$, we have $|z_0\rangle=|z\rangle, P_{i_j}|z_{j-1}\rangle=|z_j\rangle$ for $j = 1,2,\dots,q$. 
As it was shown in Ref.~\cite{GuptaBarashHen2020},
upon adding an item to or removing an item from the input list of an already evaluated divided differences
$e^{-\beta[E_0,\dots,E_q]}$, the re-evaluation can be done with only $O(s q)$ floating point operations and $O(s q)$ bytes of memory, 
where $[E_0,\dots,E_q]$ are the inputs and $s \propto \max_{i,j} |\beta E_i - \beta E_j|$.

Following the analysis of the previous subsection, to ensure the ergodicity of the Markov chain Monte Carlo,
we find that employing the following QMC updates suffices.

\subsubsection{Simple (local) swap}
Simple swap is an update that consists of selecting a random integer $m$ from $\{1,\dots,q-1\}$,
and then attempting to swap the permutation operators $P_{i_m}$ and $P_{i_{m+1}}$ in $S_{{\bf{i}}_q}$, namely:
\beq
P_{i_m} P_{i_{m+1}} \to P_{i_{m+1}} P_{i_m} \,.
\eeq
The updated sequence also evaluates to the identity.
Since the internal classical state $\vert z_{m} \rangle$ may change, implementing the swap 
involves adding a new energy $E_{z'_m}$ and removing the old one $E_{z_m}$ from the energy multi-set. 
The acceptance probability for the update satisfying the detailed balance condition is
\beq\label{eq:Pmet}
p = \min \left( 1,\frac{W_{{\cal C}'}}{W_{{\cal C}}} \right)\,,
\eeq
where it follows from Eq.~(\ref{eq:gbw}) that
\hbox{$W_{\cal C} = \Re[D_{{\cal C}}] e^{-\beta [E_{{\cal C}}]}$},
$W_{{\cal C}'} = \Re[D_{{\cal C}'}] e^{-\beta [E_{{\cal C}'}]}$,
and the energy multi-set $E_{{\cal C}'}=E_{{\cal C}} + \{E_{z'_m}\} - \{E_{z_m}\}$. 
Here, $\e^{-\beta [E_{{\cal C}}]}$ is a shorthand for $\e^{-\beta [E_{z_0},E_{z_1},\ldots,E_{z_q}]}$ for the configuration ${\cal C}$ 
and likewise for ${\cal C}'$.

\subsubsection{Pair insertion and deletion}
Pair insertion is an update consisting of selecting random integers $m \in \{1,\dots,q+1\}$ and $j \in \{1,\dots,M\}$,
and then attempting an insertion of the pair of permutation operators $P_j$ into the $S_{{\bf{i}}_q}$ sequence in such a way that
the updated operators $P_{i_m}$ and $P_{i_{m+1}}$ are $P_j$. As a result, the length of the sequences increases:  $q \to q+2$.

The updated sequence also evaluates to the identity.
The input multi-set of energies of the new configuration differs from that of the current configuration by
having two additional energies $E_{z'_m}$ and $E_{z'_{m+1}}$.
The weight of the new configuration is then $W_{{\cal C}'} = \Re[D_{{\cal C}'}] e^{-\beta [E_{{\cal C}'}]}$,
where the energy multi-set $E_{{\cal C}'}=E_{{\cal C}} + \{E_{z'_m}\} + \{E_{z'_{m+1}}\}$. 
The acceptance probability is as in Eq.~(\ref{eq:Pmet}) with the aforementioned $E_{{\cal C}'}$.

The reverse update of pair insertion is that of pair deletion.
This update can be implemented only for $q\geq 2$.
It consists of selecting a random integer $m \in \{1,\dots,q-1\}$ checking the validity of the condition
$P_{i_m} = P_{i_{m+1}}$, and attempting the removal of the pair of operators $P_{i_m}$ and $P_{i_{m+1}}$ if this condition is true.
The updated sequence also evaluates to the identity.
The input multi-set of energies of the new configuration differs from that of the current configuration by
having one fewer $E_{z_m}$ value and one fewer $E_{z_{m+1}}$ value.
The weight of the new configuration is then proportional to $e^{-\beta [E_{{\cal C}'}]}$,
where $E_{{\cal C}'}=E_{{\cal C}} - \{E_{z_m}\} - \{E_{z_{m+1}}\}$. 
To maintain the detailed balance condition, the acceptance probability turns out to be
\beq
\label{eq:Pdeletion}
p = \frac{1}{M}\min \left( 1,\frac{W_{{\cal C}'}}{W_{{\cal C}}} \right)\,.
\eeq

\subsubsection{Block swap} \label{sec:blockswap}
Block swap is an update that involves a change of the classical state $z$. Here, a random position $k\in \{1,\dots,q-1\}$ in the product $S_{{\bf{i}}_q}$ is picked such that the product is split into two (non-empty) sub-sequences, $S_{{\bf{i}}_q}=S_2 S_1$, with $S_1 = P_{i_k} \cdots P_{i_{1}}$ and $S_2 = P_{i_{q}} \cdots P_{i_{k+1}}$.  The classical state $\ket{z'}$ at position $k$ in the product is given by
\beq
|z'\rangle=S_1|z\rangle=P_{i_k} \cdots P_{i_1}|z\rangle \,,
\eeq
where $\ket{z}$ is the classical state of the current configuration.  The state $\ket{z'}$ has energy $E_{z'}$, and the state $\ket{z}$ has energy $E_{z}$.  The new block-swapped configuration is ${\cal C}'=\{|z'\rangle, S_1 S_2\}$.
The  input multi-set of energies of this configuration differs from that of the current configuration by having one fewer $E_{z}$ value and one additional $E_{z'}$ value.
The weight of the new configuration is then proportional to $e^{-\beta [E_{{\cal C}'}]}$
where the energy multi-set $E_{{\cal C}'}=E_{{\cal C}} + \{E_{z'}\} - \{E_{z}\}$. 
The acceptance probability is as in Eq.~(\ref{eq:Pmet}) with the aforementioned $E_{{\cal C}'}$.

\subsubsection{Classical updates}
\label{sec:classical_updates}

Classical moves are moves that involve a manipulation of the classical state $|z\rangle$ while leaving $S_{{\bf{i}}_q}$ unchanged.
In a single bit-flip classical move, a spin from the classical bit-string state $\ket{z}$ of ${\cal C}$ is picked at random and is subsequently flipped, 
generating a state $\ket{z'}$ and hence a new configuration ${\cal C}'$.
Calculating the weight of  ${\cal C}'$ requires the calculation of the new energy multi-set $E_{{\cal C}'}$ and recalculation of the divided differences, so it
can become computationally intensive if $q$ is large.
Classical moves should therefore be attempted with relatively low probabilities if $q$ is large.
Simply enough, the acceptance probability for a classical move satisfying the detailed balance condition is~(\ref{eq:Pmet}).

In the absence of a quantum part of the Hamiltonian ($M=0$), not only are classical moves the only moves necessary, but they are also the only moves that have nonzero acceptance probabilities. Since the initial configuration of the QMC algorithm is a random classical configuration $|z\rangle$ and an empty operator sequence $S_0=\mathbb{1}$, for a purely classical Hamiltonian, the algorithm automatically reduces to a classical thermal algorithm keeping the size of the imaginary-time dimension at zero ($q=0$) for the duration of the simulation. 

\subsubsection{Fundamental cycle completion}
Fundamental cycle completion is an update that consists of choosing a subsequence $S$ from $S_{{\bf{i}}_q}$,
choosing a fundamental cycle containing all operators of the subsequence $S$, and
attempting to replace the subsequence $S$ with the remaining operators from the selected cycle.

As is discussed in Sec~\ref{sec:ergodicity}, for the ergodicity condition to be fulfilled,
it is necessary to be able to perform the insertion (or, equivalently, completion) 
of any of the fundamental cycles.
For the probability of accepting an update to be non-negligible it is preferable that the change in the value of $q$ will be minimal. The above implies that to complete a fundamental cycle of length $l$ it is most advantageous
to replace a subsequence of length $r$ from $S_{{\bf{i}}_q}$ by the remaining $l-r$ permutation operators, where $r \approx l/2$.

We find that in some cases, there are fundamental cycles such that a simple cycle completion update,
where the subsequence $S$ comprises consecutive elements of $S_{{{\bf{i}}_q}}$, is always rejected 
for these cycles due to the zero weight of the resulting configuration [see Eq.~(\ref{eq:gbw})].
Thus, the fundamental cycle completion routine may never accept some of the fundamental cycles during the Markov process if inserted `as is'. To resolve this issue, we have developed a subroutine that we refer to as `cycle completion with gaps', which does not require the elements of the sequence $S$ to be consecutive within $S_{{\bf{i}}_q}$. Specific details of this protocol can be found in Appendix~\ref{appendix:cycle_completion}. 

\subsubsection{Composite update}

The role of the composite update is to ensure that non-fundamental cycles have the chance of being incorporated into the sequence of operators directly rather than via the concatenation of  fundamental cycles which are inserted through the cycle completion move. The composite update is required for situations where fundamental cycle insertions may have zero weight [due to the vanishing of one or more of the matrix elements in the product $D_{(z,S_{{\bf{i}}_q})}$, see Eq.~(\ref{eq:gbw})], whereas non-fundamental cycles may not.

The update consists of a combination of several basic updates and is described as follows.
(i) Perform one of the basic QMC updates equally likely: either a simple swap, a pair insertion, a pair deletion, or a fundamental cycle completion.
(ii) 
If the resulting weight is zero, reject the entire update with a probability of 1/2, and with the remaining probability, return to step (i).
(iii) Finalize the update ${\cal C}_1 \to {\cal C}_2 \to \dots \to {\cal C}_k$ with the following acceptance probability,
which satisfies the detailed balance condition:
\begin{multline}
P_{accept} \left({\cal C}_1 \to \dots \to {\cal C}_k\right)=\\
\min\left(1,
\frac{W_{{\cal C}_k}}{W_{{\cal C}_1}} R({\cal C}_1,{\cal C}_2) \dots R({\cal C}_{k-1},{\cal C}_k)\right) \,.
\end{multline}
Here, $R({\cal C},{\cal C}') = 1$ when the update ${\cal C} \to {\cal C}'$ is a simple swap, $R({\cal C},{\cal C}') = M$ when it is a pair insertion, $R({\cal C},{\cal C}') = 1/M$ when it is a pair deletion,
and $R({\cal C},{\cal C}') = (p_r(q') \cdot n_c \cdot r'!) / (p_r(q) \cdot n'_c \cdot r!)$ when it is a cycle completion (see Appendix~\ref{appendix:cycle_completion}).


To prove that the QMC updates ensure ergodicity in the entire configuration space, consider two arbitrary configurations ${\cal C} = \{ |z\rangle, S_{{\bf{i}}_q} \}$ and ${\cal C'} = \{ |z'\rangle, S_{{\bf{i'}}_{q'}} \}$
such that $W_{\cal C} \ne 0$ and $W_{\cal C'} \ne 0$.
It follows from Sec.~\ref{sec:ergodicity} that the above QMC updates allow in particular the following sequence of transformations:
${\cal C} \to {\cal C}_0 \to {\cal C}'_0 \to {\cal C}'$, where
${\cal C}_0 = \{ |z\rangle, \mathbb{1} \}$ and ${\cal C}'_0 = \{ |z'\rangle,\mathbb{1}\}$.
Hence, the transformation from ${\cal C}$ to ${\cal C}'$ is possible, and the ergodicity holds.

\subsubsection{Worm update}

An alternative to the composite update, also capable of incorporating non-fundamental cycles is a worm-type global update for PMR-QMC.
This update involves introducing a `disturbance' (or a `worm head') into the sequence of operators $S_{{{\bf i}}_q}$
by either appending $S_{{{\bf i}}_q}$ with a single operator or removing one from it
(we will refer to this addition or removal of an operator as `single operator moves').
Insertion or removal of a single permutation operator causes the sequence to evaluate to a non-identity permutation,
thus resulting in a zero-weight configuration. Consequently, the disturbed sequence must be `healed' back to 
an identity-forming sequence. The healing process involves introducing further moves, either by employing
the basic updates described above such as simple swap, fundamental cycle completion, pair insertion, and pair deletion
or by applying additional single operator moves. These single operator moves have the power to heal the sequence. 
After each such move, the instantaneous sequence $S_{{\mathbf{i}}_q}$ is checked to determine if it evaluates to the identity.
If it does, the worm update ends; if not, additional moves are required.

To make sure that detailed balance is conserved, and that the acceptance rates of intermediate worm moves are sufficiently high,
we assign non-identity intermediate configurations (sequences of operators that do not evaluate to the identity) their `natural' weight $W_{\cal C}$ 
as per Eq.~(\ref{eq:gbw}). This allows intermediate moves to be accepted or rejected with probabilities obeying detailed balance.
Additionally, to prevent the worm from straying too far from being healed (in other words, to ensure that the sequence of operators
is not too far from the identity), we introduce a small probability $p_f$ to reject the entire worm update at each intermediate state.

The worm update is specified as follows. 
(i) Start with a sequence of operators $S_{{\mathbf{i}}_q}$ that evaluates to the identity, and store  $S_{{\mathbf{i}}_q}$. 
(ii) Generate a modified sequence of operators by applying one of the following updates with equal probabilities:
either a local swap, fundamental cycle completion, pair insertion or deletion, or a single operator move.
Accept or reject the new configuration with probabilities obeying detailed balance.
(iii) The worm update ends if the new $S_{{\mathbf{i}}_q}$ evaluates to the identity.
If it does not evaluate to the identity, revert to the stored $S_{{\mathbf{i}}_q}$ and exit with probability $p_f$;
with the remaining probability, return to step (ii).

\section{Measurements}\label{sec:measurement}

The PMR formulation allows one to measure a wide range of static operators and additional dynamical quantities~\cite{pmrAdvanced}. The key to being able to do so is to write for any given operator $A$ its thermal average as
\beq
\langle A\rangle = \frac{\tr[A\e^{-\beta H}]}{\tr[\e^{-\beta H}]}=\frac{\sum_{\cal C} A_{\cal C}w_{\cal C}}{\sum_{\cal C}w_{\cal C} }.
\label{eq:meanA}
\eeq
Although, generally, both $w_{\cal C}$ and $A_{\cal C}$ are complex-valued,
both the sums $\sum_{\cal C} A_{\cal C}w_{\cal C}$ and $\sum_{\cal C}w_{\cal C}$ are real-valued
since both $H$ and $A$ are Hermitian operators. Therefore, we have
\beq \label{eq:A}
\langle A\rangle = 
\frac{\sum_{\cal C} \Re[A_{\cal C}w_{\cal C}]/\Re[w_{\cal C}]\cdot W_{\cal C}}{\sum_{\cal C} W_{\cal C} },
\eeq
where $W_{\cal C} = \Re[w_{\cal C}]$.
The quantity $\Re[A_{\cal C} w_{\cal C}]/\Re[w_{\cal C}]$
is therefore the instantaneous quantity associated with the configuration  ${\cal C} = {(z,{S_{{\bf{i}}_q}})}$
that will be gathered during the simulation.

\subsection{Measurements of standard observables}

We next provide the instantaneous quantities to be collected throughout the simulation for the following operators (i) the Hamiltonian $H$, (ii) the Hamiltonian squared $H^2$, (iii) the diagonal component of the Hamiltonian $H_\textrm{diag}=D_0$, (iv) the diagonal component squared $H_\textrm{diag}^2$, (v) the off-diagonal component of the Hamiltonian \hbox{$H_\textrm{offdiag}=\sum_{j=1}^M D_j P_j$} and (vi) the off-diagonal component of the Hamiltonian squared $H_\textrm{offdiag}^2$.

We have
$
-\partial e^{-\beta [E_{z_0},\dots,E_{z_q}]}/\partial\beta = h[E_{z_0},\dots,E_{z_q}],
$
where $h(E) = E\cdot e^{-\beta E}$.
Using the Leibniz rule for divided differences~\cite{dd:67,deboor:05}, we obtain
\be
-\frac{\partial w_{\cal C}}{\partial\beta} = H_{\cal C} w_{\cal C},
\ee
where
\be
H_{\cal C} =
	\left\{\begin{array}{ll}
        E_{z_0}, & \text{for } q = 0,\\
        E_{z_0} + \frac{e^{-\beta[E_{z_1},\dots,E_{z_{q}}]}}{e^{-\beta[E_{z_0},\dots,E_{z_q}]}}, & \text{for } q>0.
        \end{array}\right.
\ee

Therefore,
\be
\langle H\rangle = -\frac{1}{Z}\frac{\partial Z}{\partial\beta} = \frac{\sum_{\cal C} H_{\cal C} w_{\cal C} }{\sum_{\cal C}w_{\cal C} },
\ee
which coincides with Eq.~(\ref{eq:meanA}) for $A = H$.
Similarly, we have $\langle H^2\rangle = \langle H\rangle^2 - {\partial}\langle H\rangle/{\partial\beta}$, where
\be
\frac{\partial}{\partial\beta}\langle H\rangle = \frac{1}{Z}\frac{\partial}{\partial\beta}\left(\sum_{\cal C} H_{\cal C} w_{\cal C}\right) + \langle H\rangle^2.
\ee
It follows that $\left(H^2\right)_{\cal C} = (H_{\cal C})^2 - \partial H_{\cal C}/\partial\beta$, and
\be
\left(H^2\right)_{\cal C} =
	\left\{\begin{array}{ll}
        E_{z_0}^2, &\text{for } q = 0,\\
        E_{z_0}^2+\frac{\left(E_{z_0}+E_{z_{1}}\right)e^{-\beta E_{z_{1}}}}{e^{-\beta[E_{z_0},E_{z_1}]}}, &\text{for } q=1,\\
        E_{z_0}^2+\\
	(E_{z_0}+E_{z_1})\frac{e^{-\beta[E_{z_1},\dots,E_{z_q}]}}{e^{-\beta[E_{z_0},\dots,E_{z_q}]}}+& \\
	\qquad\qquad\qquad\frac{e^{-\beta[E_{z_2},\dots,E_{z_q}]}}{e^{-\beta[E_{z_0},\dots,E_{z_q}]}}, &\text{for } q>1,
        \end{array}\right.
\ee
It is straightforward to obtain the remaining expressions:
\ba
&&\left(H_\textrm{diag}\right)_{\cal C} = E_{z_0},\\
&&\left(H_\textrm{diag}^2\right)_{\cal C} = E_{z_0}^2,\\
&&\left(H_\textrm{offdiag}\right)_{\cal C} = H_{\cal C} - E_{z_0},\\
&&\left(H_\textrm{offdiag}^2\right)_{\cal C} = H^2_{\cal C} + E_{z_0}\left(E_{z_0} - 2 H_{\cal C}\right).
\ea

\subsection{Measurements of custom static operators}

We next consider the measurement of a general static operator $A$. We proceed by casting it in PMR form, i.e., as $A=\sum_i \tilde A_i \tilde P_i$ where each $\tilde A_i$ is diagonal and the $\tilde P_i$'s are permutation operators. 
The operators $\tilde{P}_i$ belong to the group $G_X$ containing all possible Pauli-$X$ tensor products, whose elements appear in the PMR representation of the Hamiltonian (see section \ref{sec:Hamiltonians}).
We note that in the most general case, the $\tilde P_i$ operators in the representation of $A$ may not all appear in the Hamiltonian's PMR decomposition Eq.~(\ref{eq:basic}). Nonetheless, we can always write
\beq \label{eq:AiPi}
\langle A \rangle = \frac{\tr[A \e^{-\beta H}]}{\tr[\e^{-\beta H}]}= \sum_i \frac{\tr[\tilde A_i \tilde P_i \e^{-\beta H}]}{\tr[\e^{-\beta H}]} \,,
\eeq
and we may focus on a single $\tilde A \tilde P$ term at a time. 
Carrying out the off-diagonal expansion~\cite{pmrAdvanced}, we end up with:
 \bea \label{eq:gen1}
&&\tr[\tilde{A} \tilde{P} \e^{-\beta H}]= \sum_z \tilde{A}(z)\sum_{q=0}^{\infty}  \sum_{{S_{{\bf{i}}_q}}}  D_{(z,S_{{{\bf i}_q}})}  \nonumber\\
&\times& {\e^{-\beta [E_{z_0},\ldots,E_{z_q}]}} \langle z| \tilde{P} S_{{{\bf i}_q}} \ket{z} \,. 
\eea
where $D_{(z,{S}_{{{\bf i}_q}})}  {\e^{-\beta [E_{z_0},\ldots,E_{z_q}]}}$ is the weight of the configuration $(z,{S}_{{{\bf i}_q}})$.

We differentiate between two cases:
(i) the operator $\tilde P$ appears in the Hamiltonian or can be written as a product of permutation operators that appear in the Hamiltonian,
(ii) The operator $\tilde P$ does not appear in the Hamiltonian and cannot be written as a product of permutation operators that do.

In the latter case,
since every $S_{{{\bf i}_q}}$ sequence consists of the permutation operators that appear in the Hamiltonian,
we always have $\langle z | \tilde P S_{{{\bf i}_q}} | z \rangle = 0 $. Then, it follows 
from Eq.~(\ref{eq:gen1}) that $\langle \tilde A \tilde P \rangle = 0$.

In the former case, the operator to be measured has the form $A=\tilde{A} \tilde{P}$ where $\tilde{A}$ is diagonal and 
$\tilde{P} = P_{i_1} P_{i_2} \cdots P_{i_k}$.
We modify Eq.~(\ref{eq:gen1}) so that $(z,\tilde{S}_{{{\bf i}_q}})$ with  $\tilde{S}_{{{\bf i}_q}} = \tilde{P} S_{{{\bf i}_q}}$ is seen as a configuration instead of $(z,{S}_{{{\bf i}_q}})$.
Thus, we arrive at: 
\beq
\langle A \rangle = 
\frac
{
\sum_{(z,{\tilde{S}_{{\bf{i}}_q}})} w_{(z,{{\tilde{S}_{{\bf{i}}_q}}})} 
 M_{\tilde A \tilde P}{(z,{\tilde{S}_{{\bf{i}}_q}})}  
}
{ \sum_{(z,{\tilde{S}_{{\bf{i}}_q}})} w_{(z,{{\tilde{S}_{{\bf{i}}_q}}})}},
\label{eq:customOperator1}
\eeq
where
\beq
M_{\tilde{A}\tilde{P}}{(z,{\tilde{S}_{{\bf{i}}_q}})} = \delta_{\tilde{P}} \tilde{A}(z) \frac{1}{D_{(z,\tilde{P})}} 
\frac{  {\e^{-\beta [E_{z_0},\ldots,E_{z_{q-k}}]}} } { {\e^{-\beta [E_{z_0},\ldots,E_{z_{q}}]}} } \,.
\label{eq:customOperator2}
\eeq
In the above, $\delta_{\tilde{P}}=1$ if the leftmost operators of $\tilde{S}_{{{\bf i}_q}}$ are $P_{i_1} P_{i_2} \cdots P_{i_k}$ and is zero otherwise,
and 
\beq
D_{(z,\tilde{P})} = \frac{D_{(z,\tilde{S}_{{{\bf i}_q}})}}{D_{(z,{S}_{{{\bf i}_q}})}} = \prod_{m=1}^{k} \langle z_{q-m+1} |D_{i_m}| z_{q-m+1} \rangle.
\eeq

The above formulation allows automatic measurements of static operators as long as these are set up as linear combinations of Pauli strings as in Eq.~(\ref{eq:PauliStrings}).
For an arbitrary static operator $A$ given as a linear combination of Pauli strings, one is required to cast the expression in the form $\sum_i \tilde A_i \tilde P_i$.
As a next step, the $\tilde P_i$ operators are expressed in terms of the permutations $P_i$ of the Hamiltonian. Here, the Gaussian elimination over mod-$2$ addition can be used~\cite{GaussianElimination1991}. Finally, Eqs.~(\ref{eq:customOperator1}), (\ref{eq:customOperator2}) are employed to compute the thermal average $\langle A\rangle$.

However, for the thermal average of a  custom observable to be computed correctly, it is necessary to account for the following restrictions on its structure.
We find that employing Eqs.~(\ref{eq:customOperator1}) and (\ref{eq:customOperator2})
lead to obtaining the correct value of the thermal average $\langle A\rangle$ under the  condition that $\tilde{A}(z) = 0$ for all basis states $|z\rangle$ for which $D_{(z,\tilde{P})} = 0$. Otherwise, it is possible that the following conditions hold for some of the configurations:
$D_{(z,\tilde{P})} = 0$ and $D_{(z,{\tilde S}_{{{\bf i}_q}})} = 0$ while $D_{(z,{S}_{{{\bf i}_q}})} \ne 0$ and $\tilde A(z) \ne 0$,
where $\tilde{S}_{{{\bf i}_q}} = \tilde{P} S_{{{\bf i}_q}}$.
Such configurations $(z,{S}_{{{\bf i}_q}})$ have a nonzero contribution in Eq.~(\ref{eq:gen1}),
but the Markov chain does not generate the corresponding configurations $(z,{\tilde{S}}_{{{\bf i}_q}})$
because they have zero weights: $W_{(z,{{\tilde{S}_{{\bf{i}}_q}}})} = 0$.

In particular, it is possible to obtain the thermal average of any observable of the form $A = \sum_i A_i D_i P_i$, where each $A_i$ is an arbitrary diagonal matrix and each $D_i P_i$ is a generalized permutation matrix appearing in Eq.~(\ref{eq:basic}). Any operator $A$ satisfying $\langle z | A | z' \rangle = 0$ for all basis states $\ket{z}$ and $\ket{z'}$ such that $\langle z | H | z' \rangle = 0$, can be written in such a form and hence its thermal average~(\ref{eq:customOperator1}) will be computed correctly.

In addition to the above measurement protocol, there is also a way to
correctly obtain a thermal average of a static operator $A$ 
which can be written in the form $A = \sum_i \tilde{P}^{(i)}$, where each $\tilde{P}^{(i)}$ is a matrix of the general form
\beq
\tilde{P}^{(i)} = A_k^{(i)} D_{j_k} P_{j_k} \dots A_2^{(i)} D_{j_2} P_{j_2} \cdot A_1^{(i)} D_{j_1} P_{j_1} \cdot A_0^{(i)},
\eeq
where $A_0^{(i)}, \dots, A_k^{(i)}$ are arbitrary diagonal matrices and
$D_{j_1} P_{j_1}, \dots, D_{j_k} P_{j_k}$ are the generalized permutation matrices
appearing in Eq.~(\ref{eq:basic}).


Focusing on a single such term,
we calculate the expectation value for an observable of the form
\beq\label{eq:pbar}
\bar{P} = A_k D_{j_k} P_{j_k}  \cdots A_2 D_{j_2} P_{j_2} A_{1} D_{j_1} P_{j_1} A_{0}\,,
\eeq 
where each of the $D_{j_k} P_{j_k}$ appear in the PMR decomposition of the Hamiltonian and the $A_k$ matrices are arbitrary diagonal operators. 
Carrying out the off-diagonal expansion for $\tr[\bar{P}  \e^{-\beta H}]$, we first obtain
\begin{multline}
\label{eq:AP}
\tr[ \bar{P} \e^{-\beta H}]
= \sum_z M_{\bar{P}}{(z,{{{S}_{{{\bf i}_q}}}})}   \sum_{q=0}^{\infty}  \sum_{{{{S}_{{{\bf i}_q}}}}}  D_{(z,S_{{{\bf i}_q}})} 
\times\\ \times  {\e^{-\beta [E_{z_0},\ldots,E_{z_q}]}} \langle z|\tilde{P} S_{{{\bf i}_q}} \ket{z} \,,
\end{multline}
where
\beq
M_{\bar{P}}{(z,{{{S}_{{{\bf i}_q}}}})} = A_k(z_{q+k}) d_{z_{q+k}}^{(j_k)} \cdots    A_1(z_{q+1}) d_{z_{q+1}}^{(j_1)}  A_{0}(z_{q}) \,,
\eeq
with $A_i(z_j) \equiv \langle z_j| A_i |z_j\rangle$.
Note that there is a one-to-one correspondence between non-vanishing terms $\langle z|\tilde{P} S_{{{\bf i}_q}} \ket{z} =\langle z|P_{j_k}  \cdots P_{j_2}  P_{j_1}S_{{{\bf i}_q}} \ket{z}$ and non-vanishing terms $\langle z|S_{{{\bf i}_q}} \ket{z}$ which appear in the partition function expansion.
We can therefore rewrite the above as:
\begin{multline}
\hspace{-0.2cm}
\tr[ \bar{P} \e^{-\beta H}]= \sum_z \left(\prod_{j=0}^{k} A_j (z_{q+j})\right) \sum_{q=0}^{\infty}  \sum_{S_{{{\bf i}_q}}}  D_{(z,S_{{{\bf \tilde{i}}}})}  
\times \\ \times
{\e^{-\beta [E_{z_0},\ldots,E_{z_q}]}}
\frac{D_{(z,S_{{{\bf i}}})}  {\e^{-\beta [E_{z_0},\ldots,E_{z_{q+k}}]}}}{D_{(z,S_{{{\bf \tilde{i}}}})}  {\e^{-\beta [E_{z_0},\ldots,E_{z_{q+k}}]}}} \langle z|S_{{{\bf \tilde{i}}}} \ket{z} \,. 
\end{multline}
where $D_{(z,S_{{{\bf \tilde{i}}}})}  {\e^{-\beta [E_{z_0},\ldots,E_{z_{q+k}}]}}$ is the weight of configuration $(z,S_{{{\bf \tilde{i}}}})$ with  $S_{{{\bf \tilde{i}}}} =P_{j_k}  \cdots P_{j_2}  P_{j_1} S_{{{\bf i}_q}} $. 
This gives:
\begin{multline}
\label{eq:z3}
\tr[\bar{P} \e^{-\beta H}]= \sum_z \sum_{q=0}^{\infty}  \sum_{S_{{{\bf i}_q}}}  D_{(z,S_{{{\bf i}_q}})}  {\e^{-\beta [E_{z_0},\ldots,E_{z_q}]}} 
\times \\ \times
M_{\bar{P}}{(z,S_{{{\bf i}_q}})}  \langle z|S_{{{\bf i}_q}} \ket{z} \,. 
\end{multline}
where $M_{\bar{P}}{(z,S_{{{\bf i}_q}})} $ is redefined as
\beq \label{eq:mAP}
M_{\bar{P}}{(z,S_{{{\bf i}_q}})} = \delta_{\bar{P}} \left(\prod_{j=0}^{k} A_j (z_{q-k+j})\right)  \frac{\e^{-\beta[E_{z_{0}},\ldots,E_{z_{q-k}}]}}{\e^{-\beta[E_z,\ldots,E_{z_q}]}} \,.
\eeq
In the above expression,  $\delta_{\bar{P}}=1$ if the leftmost operators of $S_{{{\bf i}_q}}$ is $P_{j_1} P_{j_2} \cdots P_{j_k}$ and is zero otherwise.

Denoting $\widetilde{A}(z,S_{{{\bf i}_q}})= \sum_{i=0}^q M_{\bar{P}_i}{(z,S_{{{\bf i}_q}})}$, we can thus write $\langle A \rangle$ as
\beq
\langle A \rangle =\left\langle \sum_i \bar{P}^{(i)}\right\rangle = 
\frac
{
\sum_{(z,S_{{{\bf i}_q}})} w_{(z,S_{{{\bf i}_q}})} 
\widetilde{A}(z,S_{{{\bf i}_q}})
}
{ \sum_{(z,S_{{{\bf i}_q}})} w_{(z,S_{{{\bf i}_q}})}} \,.
\eeq

\subsection{Monitoring the average sign}
\label{sec:sign}

A necessary condition for the proper importance sampling of partition function weights is that all weights are nonnegative. Whenever the partition function expansion produces negative weights  [see Eq.~(\ref{eq:gbw})], the system is said to possess a sign problem~\cite{vgp}. In the presence of a sign problem, configuration weights cannot be treated as un-normalized probabilities as they should in Markov chain Monte Carlo simulations. In such cases, a common workaround is to take the configuration weights to be the absolute values of the original ones $|W_{\cal C}|$~\cite{Troyer2005,GuptaHen2020,vgp}.
By doing so, a thermal average of an observable $A$ is re-written as
\beq
\langle A\rangle = \frac{\langle A\cdot\sgn(W)\rangle_{|W|}}{\langle \sgn(W)\rangle_{|W|}},
\label{eq:Amean}
\eeq
where the average sign
\beq
\langle \sgn \rangle = \langle \sgn(W)\rangle_{|W|} = \frac{\sum_{\cal C} W_{\cal C}}{\sum_{\cal C} |W_{\cal C}|} 
\eeq
is monitored throughout the simulation.
The average sign is usually considered a figure of merit of how adverse the sign problem is. For models that do not have a sign problem, all weights are positive,  $\langle\sgn\rangle = 1$ and the expression for $\langle A \rangle$ naturally reduces to its original form, Eq.~(\ref{eq:A}). For sign-problematic systems, $\langle\sgn\rangle \approx 0$, and one would expect to obtain extremely large error bars for $\langle A \rangle$ that would as a result require an exponentially long simulation time for an accurate computation of observables. 

As we demonstrate in the next section, in some cases accurate calculations are achievable even when $\langle\sgn\rangle$ is relatively small. In Appendix~\ref{appendix:errors} we provide an improved approximation of $\langle A\rangle$ in the presence of a sign problem and an approximation for the statistical error $\sigma(A)$, which can be determined during the simulation.

\begin{figure}[t]
\includegraphics[width=.98\columnwidth]{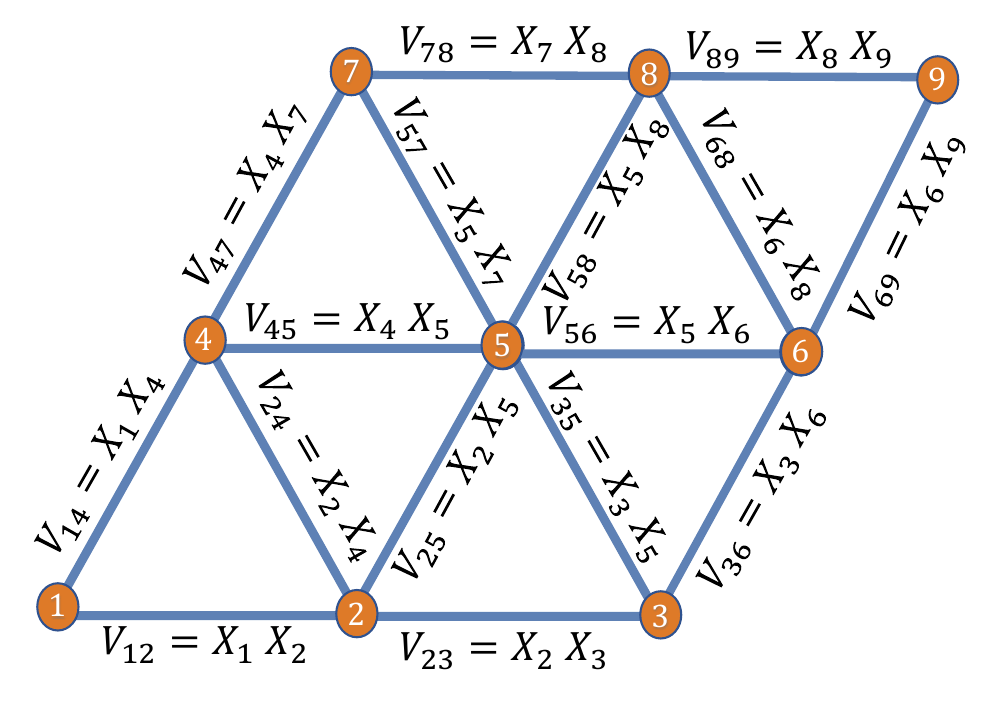}
\caption{\label{fig:triLatt} A triangular lattice with open boundary conditions. Here the side length is $L=3$. }
\end{figure}

\section{Results and discussion}\label{sec:Results}

In this section, we demonstrate the success of our method in simulating a variety of large-scale quantum many-body systems, taking as test cases models that would help highlight the extensive scope of the algorithm.

Wherever exact calculations were possible, we have verified the correctness and accuracy of our technique by ensuring that the calculated values agree with exact values. 

\subsection{The \texorpdfstring{XY}{XY} model on a triangular lattice}

\begin{figure}[htp]
\includegraphics[width=\columnwidth]{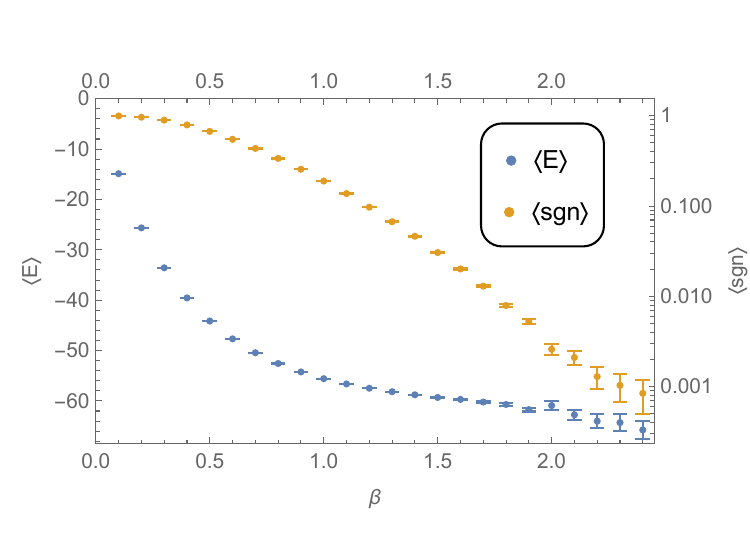}
\includegraphics[width=\columnwidth]{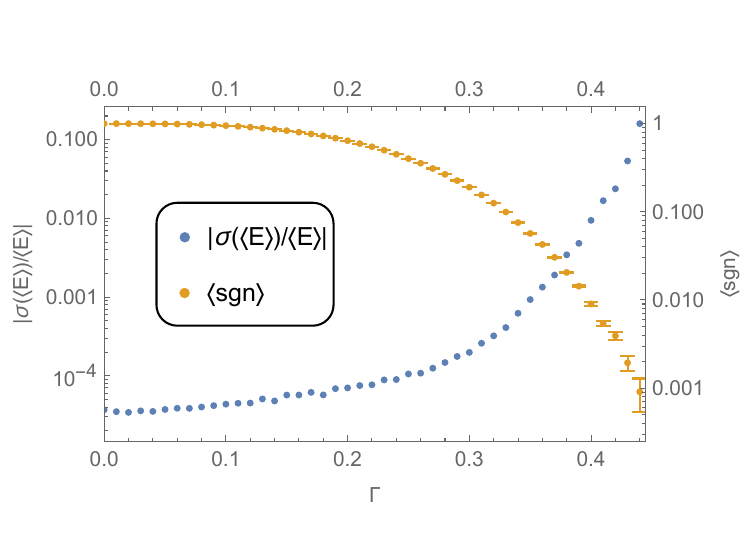}
\caption{Top: Dependence of mean energy and $\langle\sgn\rangle$ on $\beta$ for $L=8$ and $\Gamma = 0.3$.
Bottom:  $\langle\sgn\rangle$ and relative error of the mean energy as a function of $\Gamma$ for $L=8$ and $\beta = 1$.
}
\label{fig:E}
\end{figure}

We consider the prototypical quantum anisotropic XY ($XZ$) model~\cite{XY1,XY2,XY3} on a triangular lattice 
with open boundary conditions (see Fig.~\ref{fig:triLatt}). This model and variants thereof have been used 
extensively as simplified models for a variety of physical systems such as liquid helium, high-$T_c$ superconductors, 
anisotropic magnets and more. The Hamiltonian we study is
\beq \label{eq:Htri}
H=J \sum_{\langle j k \rangle} Z_j Z_k + \Gamma \sum_{\langle j k \rangle} X_j X_k \,,
\eeq
where $\langle j k \rangle$ denotes neighbors on a triangular lattice with $n=L^2$ spins containing $L$ sites on each side. The above XY model 
gives rise to a severe sign problem, preventing a true quantitative understanding of the phase diagram of the model
if $J>0$ and $\Gamma>0$, which will be our region of interest.
Specifically, we shall consider a system with $L=8$ and $J=1$ and allow the parameter $\Gamma$, which serves as the strength of the quantum component of the Hamiltonian, to vary.

The PMR form of the Hamiltonian, Eq.~(\ref{eq:Htri}), is
\beq
H = D_0 + \Gamma \sum_{\langle j k\rangle}  V_{jk} \label{eq:PMR_example}\,.
\eeq
Here, $D_0=J \sum_{\langle j k \rangle} Z_j Z_k $ is the `classical' component of the Hamiltonian that is diagonal in the computational basis. 
The set $\{V_{jk}=X_j X_k \}$ consists of off-diagonal permutation operators that give the system its  `quantum dimension' and obey ${V}_{ij} | z \rangle = | z' \rangle$
for every basis state $|z\rangle$, where $ | z' \rangle \neq |z\rangle$ is also a basis state differing from $|z\rangle$ by two spin flips. 
There are $M=3L^2-4L+1$ off-diagonal operators $V_{jk}$, one for each edge of the lattice (see Fig.~\ref{fig:triLatt}).

Since $V_{ij} V_{jk} V_{ki} = \mathbb{1}$ for every triplet of spins with indices $i, j$ and $k$ that form a basic triangle of the lattice (or a triangular plaquette), we conclude that the model admits $2(L-1)^2$ fundamental cycles, one for each basic triangle (see  Sec.~\ref{sec:fundamentalcycles}).

\begin{table}[t]
\begin{tabular}{|c|c|c|c|c|c|c|}
\hline
$\Gamma$ & $\langle\sgn\rangle$ & $\langle q\rangle$ & $\mathrm{\max}(q)$ & time (s.) per \\
         &                                   &                    &                   & MC update\\
\hline
$0.05$      &   $0.994$    &   $0.2$    &    $9$  &  $2.0\cdot 10^{-6}$ \\
$0.1$       &   $0.95$     &   $0.84$   &   $13$  &  $2.7\cdot 10^{-6}$ \\
$0.2$       &   $0.64$     &   $4$      &   $30$  &  $5.4\cdot 10^{-6}$ \\
$0.3$       &   $0.19$     &   $11$     &   $46$  &  $1.2\cdot 10^{-5}$ \\
$0.35$      &   $0.06$     &   $17$     &   $59$  &  $1.8\cdot 10^{-5}$ \\
$0.4$       &   $0.01$     &   $27$     &   $75$  &  $3.5\cdot 10^{-5}$ \\
$0.45$      &   $0.001$   &   $41$     &   $93$  &  $5.7\cdot 10^{-5}$ \\
\hline
\end{tabular}
\caption{Dependence of the expansion order and average sign on the off-diagonal strength $\Gamma$ for the antiferromagnetic XY model on a triangular lattice. Calculations are shown for $L = 8$, $J = 1$, $\beta = 1$.}
\label{tab:tab1}
\end{table}

Figure~\ref{fig:E} shows the computed mean energy, its error bar, and $\langle\sgn\rangle$ as a function of $\beta$ and $\Gamma$.
Here, we have used $2\cdot 10^9$ Monte-Carlo updates.
As is shown in Fig.~\ref{fig:E}, the observables can be accurately calculated using our method
even in cases where the values of $\langle\sgn\rangle$ are as small as $10^{-3}$.
However, as expected, the values of $\langle\sgn\rangle$ decrease exponentially with both $\beta$ and $\Gamma$.

Table~\ref{tab:tab1} shows the increase in  complexity of the calculation with the severity of the sign problem
in more detail. In particular, one can see that the wall-clock time is roughly proportional to $\langle q\rangle$ for $\langle q\rangle >1$ (in agreement with prior results pertaining to divided-difference calculations~\cite{GuptaBarashHen2020}). This is because most of the computing time is spent on the calculation and reevaluation of the divided differences.

\subsection{The \texorpdfstring{XY}{XY} model on a square lattice: Dependence of convergence time on temperature}

\begin{figure}[th]
\centering
\includegraphics[width=\columnwidth]{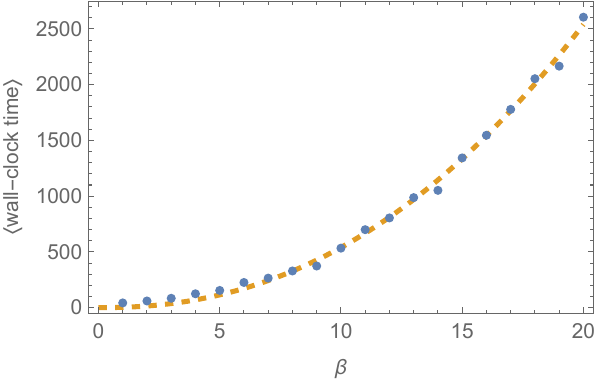}%
\llap{\makebox[0.82\columnwidth][l]{\raisebox{0.29\columnwidth}{\includegraphics[width=0.5\columnwidth]{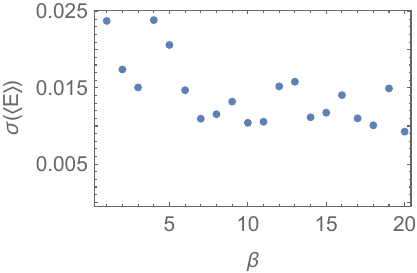}}}}
\caption{Calculations of the XY-model on a square lattice with periodic boundary conditions:
wall-clock time in seconds of $10^7$ QMC updates as a function of $\beta$.
Parameters: $L=8$, $J=1$, $\Gamma=-0.25$. The dashed line is $3.14\times \beta^{2.23}$.
Inset: estimated statistical error of $10^7$ QMC updates as a function of $\beta$. 
}
\label{fig:periodic}
\end{figure}


We next investigate the dependence of simulation runtimes on the inverse-temperature $\beta$, using as a test case the XY-model on a square lattice with periodic boundary conditions imposed. Similar to the XY model on the triangular lattice discussed in the previous subsection, the permutations of the XY model on the square lattice are two-body $X X$ interactions. The fundamental cycles are therefore length-4 products of permutations corresponding to edges surrounding each plaquette. On top of these, the periodic boundary conditions induce additional cycles that wrap around the lattice. For an $8 \times 8$ square lattice, the nullspace consists of $65$ cycles, of which $56$ are of length four and $9$ of length eight wrapping around the lattice either horizontally or vertically. 

Figure~\ref{fig:periodic} shows the wall-clock time of $10^7$ QMC updates as a function of $\beta$, fitted by the dashed curve with the optimal fit of $3.14\times \beta^{2.23}$.
The inset shows the estimated error $\sigma(\langle E\rangle)$ of $10^7$ QMC updates versus $\beta$.
The statistical error turns out to be of the same order of magnitude across the entire tested temperature range. 

We thus find that the convergence time of the algorithm grows relatively slowly with $\beta$, obeying a modest power-law, indicating that low-temperature simulations are readily attainable.

\subsection{Topological models}

As mentioned above, in the presence of a non-trivial topology, e.g., where periodic boundary conditions are imposed or for other nonzero genus models, there may exist fundamental cycles becoming longer with the system size such as the cycles `wrapping around the system', consisting of an extensive number of permutation operators (as was the case in the XY model on a square lattice with periodic boundary conditions discussed above). 
For such models, some fundamental cycle lengths will be of the order $O(N)$, i.e., grow with system size, (as opposed to being $O(1)$. For that reason, these models are usually exceptionally difficult to study as they require so-called `global' rather than local moves. 

To demonstrate that our proposed method can successfully solve these models as well, we next present some results pertaining to the well-known `toric code' model. 
The toric code is defined on a periodic two-dimensional lattice (a torus), usually chosen to be the square lattice, with a spin-1/2 particle located on each edge. The Hamiltonian of the toric code is given by~\cite{Kitaev2003,Field2018}: 
\beq
H_{\rm toric} = J\left( \sum_v A_v  + \sum_p B_p\right) \,,
\eeq
where $J>0$ and $A_v = \prod_{i \in v} X_i$ and $B_p = \prod_{i \in p} Z_i$
with $i \in v$ denoting the edges touching the vertex $v$, and $i \in p$ denoting the edges surrounding the plaquette $p$. In PMR, the Hamiltonian is rewritten as
\beq
H_{\rm toric}  = D_0 + \sum_v D_v P_v 
\eeq
where $D_0= \sum_p B_p$, $P_v =A_v$ and $D_v=J \cdot \mathds{1}$. 
The only fundamental cycle is equal to the product of all plaquette terms $P\equiv\prod_v P_v$.

The local moves of insertion and deletion of pairs of permutation operators are however not ergodic on their own. This can be seen by noticing that the annihlation/creation local updates implies that plaquette terms $P_v$ appear in even numbers. On the other hand, the product of all plaquette terms $P\equiv\prod_v P_v$ also evaluates to the identity and so sequences where all plaquette terms are odd-numbered should also be accounted for. There are thus two topological sectors: an even one and an odd one. For the simulation to sample configurations in both, there must be a global move that jumps between the two and which changes the parity of all plaquette operators.
As follows from Sec.~\ref{sec:ergodicity}, the fundamental cycle completion accomplishes this move.

\begin{figure}[t]
\includegraphics[width=\columnwidth]{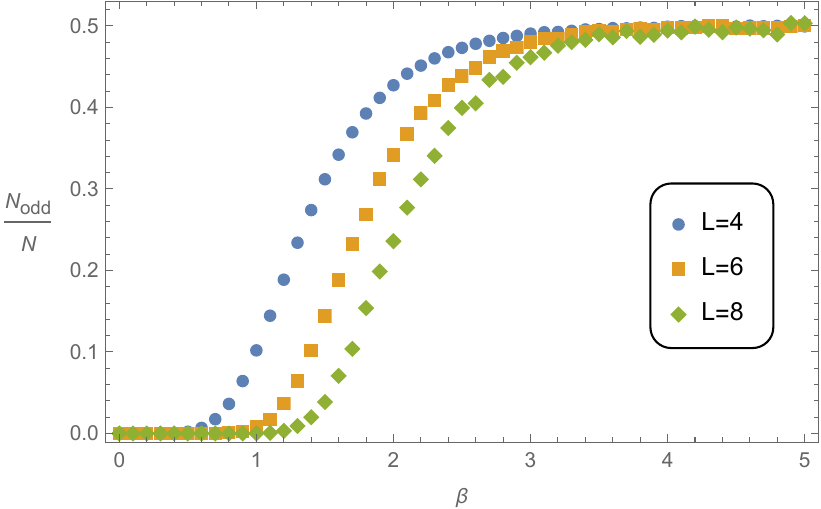}
\caption{Toric code: the fraction of time in the odd sector as a function of $\beta$ for $L=4$, $L=6$, and $L=8$.}
\label{fig:toric_code}
\end{figure}

Figure~\ref{fig:toric_code} shows $N_{\textrm{odd}}/N$ versus $\beta$, where $N_{\textrm{odd}}$ is the number of visited odd-sector configurations, and $N$ is the total number of visited configurations.
As can be seen, the Markov chain switches easily between the even sector and the odd one for sufficiently low temperatures, indicating that the Markov chain is ergodic and the algorithm works properly.
The above results also agree with the expected behavior of toric code,
where transitions between ground states are generated by pairs of excitations at sufficiently low temperatures~\cite{Freeman2014}.

\subsection{Random \texorpdfstring{spin-$1/2$}{spin-1/2} Hamiltonians}\label{sec:Results2}

\begin{figure}[t]
\includegraphics[width=\columnwidth]{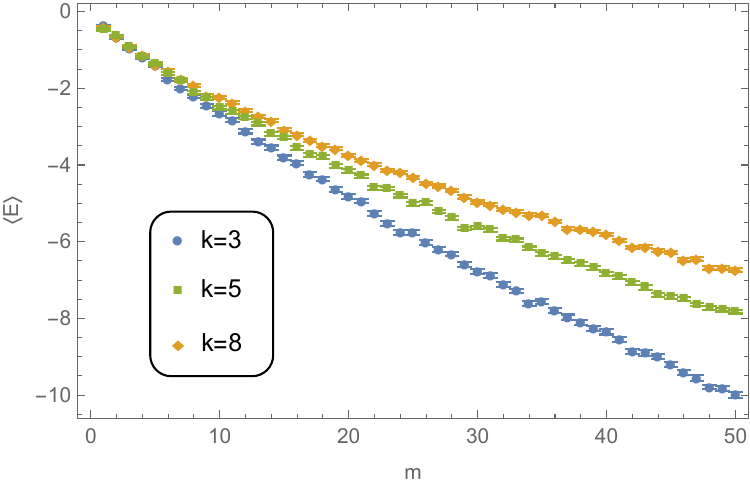}
\includegraphics[width=\columnwidth]{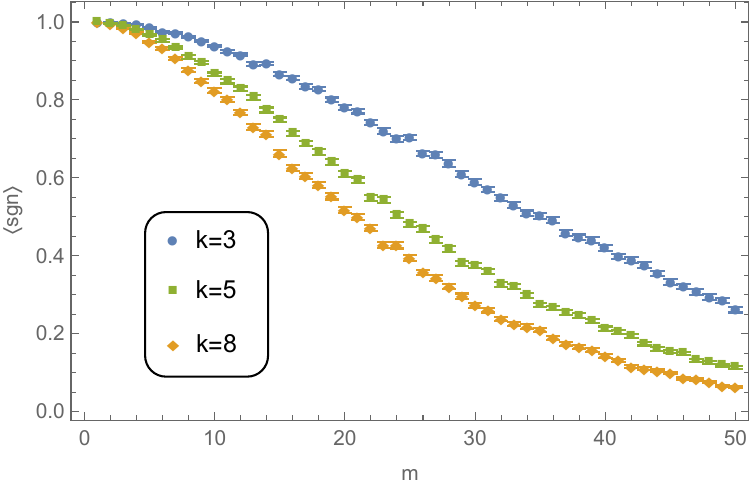}
\caption{Top: Average energy $\langle E\rangle$ over $200$ randomly generated Hamiltonian instances as a function of $m$ for random $k$-local $40$-spin  Hamiltonians for $k=3$, $k=5$, and $k=8$. Bottom: A similar plot for $\langle\sgn\rangle$, averaged over the $200$ Hamiltonian instances. Here, $\beta=1$.}
\label{fig:random_k_local_beta=1}
\end{figure}

\begin{figure}[t]
\includegraphics[width=\columnwidth]{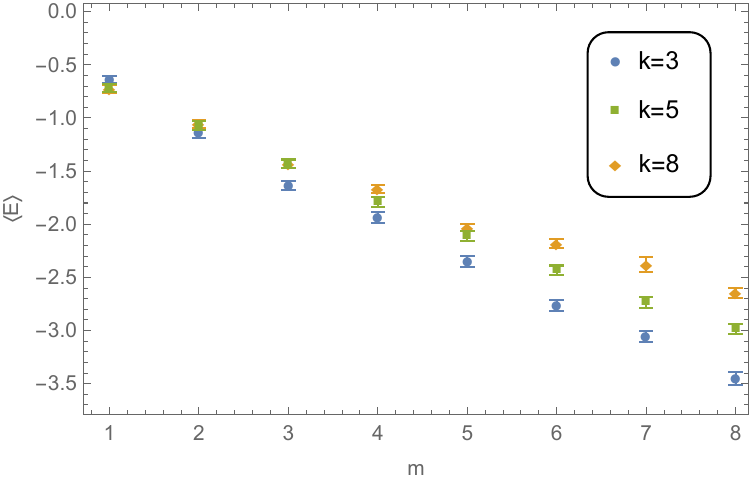}
\includegraphics[width=\columnwidth]{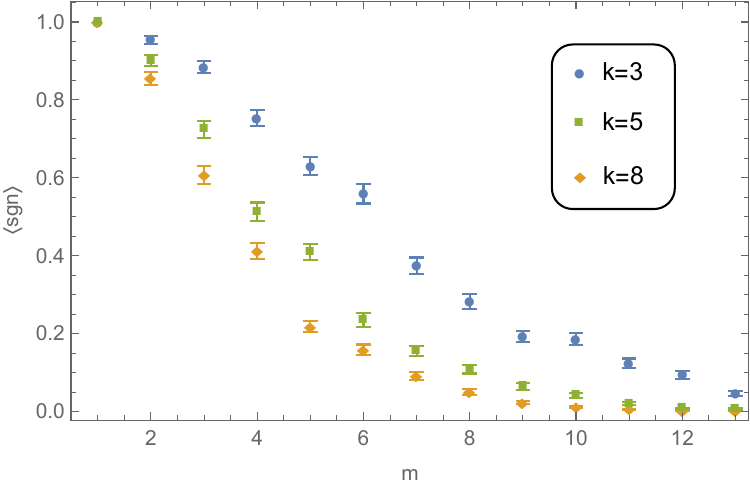}
\caption{Top: Average energy $\langle E\rangle$ over $200$ randomly generated Hamiltonian instances as a function of $m$ for random $k$-local $40$-spin  Hamiltonians for $k=3$, $k=5$, and $k=8$. Bottom: A similar plot for $\langle\sgn\rangle$, averaged over the $200$ Hamiltonian instances. Here, $\beta=5$. }
\label{fig:random_k_local_beta=5}
\end{figure}

To illustrate the versatility of our algorithm, we next present simulation results for randomly generated spin-$1/2$ Hamiltonians. We produce random $n$-spin $m$-term $k$-local spin-$1/2$ Hamiltonian by adding together $m$ randomly generated Pauli strings. To create a $k$-local Pauli string, we first sample $k$ spin indices $i_1, \ldots, i_k$ from the set of spin indices $\{1, \ldots, n\}$.  For each chosen index, we pick at random an operator from the set $\{ X, Y, Z\}$, thereby creating a Pauli string with locality $k$. Our final random Hamiltonian attains the form $\sum_i c^{(i)} S^{(i)}$, where each randomly generated Pauli string $S^{(i)}$ is multiplied by a real-valued coefficient $c^{(i)}$ randomly drawn from the interval $[-1,1]$. 

To demonstrate the ease with which our approach allows the simulation of such systems, we have generated random $40$-spin $m$-term Hamiltonians with $m$ varying from $m=1$ to $m=50$, simulating $200$ randomly generated instances per each value of $m$ at $\beta=1$ for three choices of locality $k=3$, $k=5$, and $k=8$. Figure~\ref{fig:random_k_local_beta=1} (top) shows the average of the energy $\langle E\rangle$ over the $200$ instances as a function of $m$. The error bars indicate magnitude of fluctuations of the averaged energy. The bottom panel of Fig.~\ref{fig:random_k_local_beta=1} depicts the average sign $\langle\sgn\rangle$, averaged over the $200$ instances per each choice of $m$ and $k$.
Similarly, Figure~\ref{fig:random_k_local_beta=5} depicts $\langle E\rangle$ and $\langle\sgn\rangle$ as a function of $m$ for $\beta=5$.

\subsection{Classically frustrated spin models}

It is important to note that while our approach guarantees a correct equilibrium distribution of the Markov chain for any input spin-$1/2$ Hamiltonian, a universal rapid mixing of the Markov chain cannot be ensured in general.

One class of many-body systems that is known to considerably hinder the convergence of Monte Carlo algorithms -- classical or quantum --  is that of strongly frustrated spin models. For these, there exist competing terms in the Hamiltonians the minimization of one directly conflicts with the minimization of others creating `frustration'. 
For such Hamiltonians, employing a Markov chain whose classical updates are based on single spin-flip Metropolis moves will result in slow mixing. This is because single spin-flip moves may cause the simulation to 
get trapped in metastable regions of configuration space (i.e., `local minima').

In many cases, the slowdown caused by frustration can be significantly mitigated (but not fully cured in general due to the NP-hardness of the underlying problem) by replacing the classical Metropolis updates
with suitable cluster updates if such exist~\cite{Kandel1990,Zhang1994,Coddington1994,Houdayer2001,Rakala2017}.
Replacing classical Metropolis updates (see Sec.~\ref{sec:classical_updates}) with relevant cluster updates
often results in a much faster converging algorithm.\footnote{These classical cluster moves can be generalized to quantum cluster moves for which the configuration weights are different from Boltzmann weights. However, such a generalization is beyond the scope of the present paper.}
Another efficient approach to some frustrated systems is to combine 
the QMC algorithm with parallel tempering~\cite{Swe1986,Hukushima1996} or population annealing~\cite{Machta2010,PA2017,PA2019}.

\begin{figure}[th]
\centering
\includegraphics[width=\columnwidth]{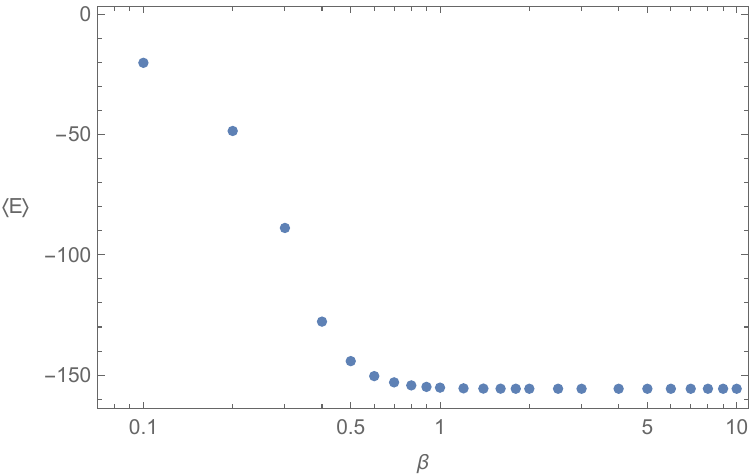}%
\llap{\makebox[0.64\columnwidth][l]{\raisebox{0.215\columnwidth}{\includegraphics[width=0.63\columnwidth]{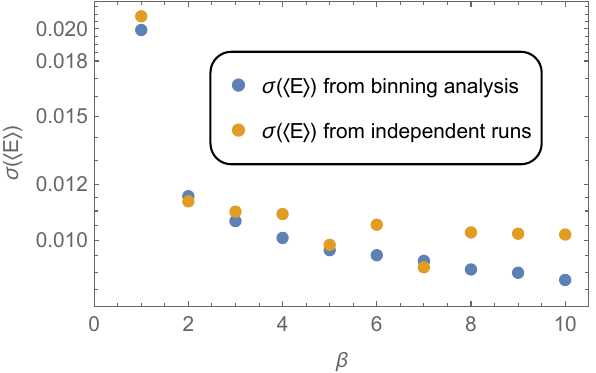}}}}
\caption{
Dependence of mean energy on $\beta$ for 
the slightly frustrated XY model on a triangular lattice.
Inset: error estimation.
Parameters: $L=8$, $\Gamma=-0.25$, $J=1$ for a few randomly selected edges and $J=-1$ for the remaining edges.
}
\label{fig:reduced_frustration}
\end{figure}

A simple way to detect whether geometric frustration leads to a computational inefficiency in a Monte Carlo calculation
is to compare the statistical error obtained from binning analysis (see Appendix~\ref{appendix:errors})
with the statistical error estimated from multiple completely independent, and hence uncorrelated, runs of the algorithm.
If the system is thermalized throughout the course of the simulation, these two estimates should roughly agree. For a highly frustrated system at a low temperature, for which the auto-correlation time is longer than the timescale of the simulation, the error obtained from the binning analysis is significantly underestimated due to the measurements being not fully decorrelated.

Figure~\ref{fig:reduced_frustration} shows the computed mean energy 
as a function of $\beta$ for a slightly frustrated model where only a small fraction of the edges are antiferromagnetic.
The inset shows the error estimates from binning analysis and from independent runs.
This example suggests that it is possible to compute the observables with high accuracy at low temperatures
if the fraction of frustrated plaquettes is sufficiently low.

\subsection{Code parallelization}

\begin{figure}[th]
\includegraphics[width=\columnwidth]{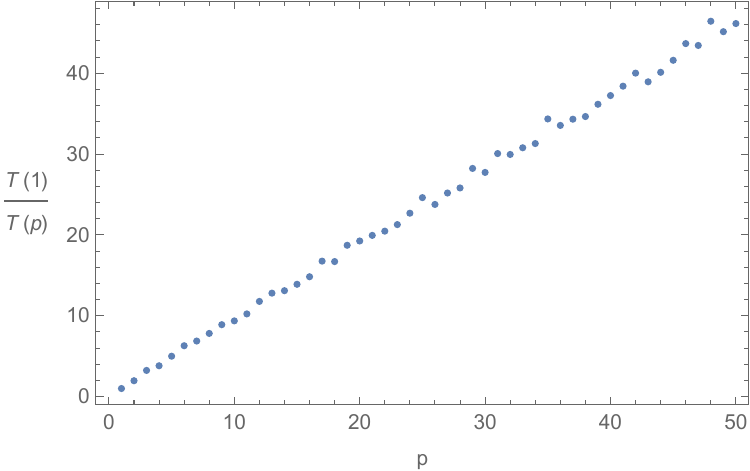}
\caption{Strong scaling speedup. 
Here, $T(p)$ is the time required to complete the same amount of work using the parallel setup with $p$ processing threads.
The total numbers of initial thermalization QMC updates and main QMC updates are equal to 
$2.5\times 10^7$ and $2.5\times 10^8$, respectively, for each of the parallel setups.
Calculations were performed for the XY-model on a square lattice with periodic boundary conditions
with $L=8$, $J=1$, $\Gamma=-0.25$.
}
\label{fig:strong_scaling}
\end{figure}

Markov chain Monte Carlo algorithms are naturally well-suited for massively parallel simulations, where independently run Markov chains contribute equally to the collection of statistics.
As shown in Fig.~\ref{fig:strong_scaling}, the algorithm, indeed, demonstrates a near-perfect strong scaling speedup, with the only restriction being that each of the parallel processes is expected to employ a sufficient number of initial thermalization (warmup) steps prior to measurement collection.

In addition to sequential execution of the C++ program code, our software package includes capabilities for executing the code in a parallel fashion on high-performance compute clusters using Message Passing Interface (MPI) protocols~\cite{mpi}, allowing for extensive parallelization of the algorithm~\cite{githubPMRQMC}.

\section{Summary and outlook}\label{sec:conclusions}
We presented a universal, parameter-free, Trotter-error-free quantum Monte Carlo scheme capable of simulating, for the first time, arbitrary spin-$1/2$ Hamiltonians. We have demonstrated that the permutation matrix representation of Hamiltonians allows one to automatically produce QMC updates that are provably ergodic and satisfy detailed balance, thereby ensuring the convergence of the Markov chain to the proper thermal equilibrium. We have in addition illustrated how a wide range of observables may be calculated throughout the simulation. 

Our algorithm therefore allows one to study the equilibrium properties of essentially any conceivable spin-$1/2$ system with a single piece of code that accepts as input a description of a Hamiltonian. This is in stark contrast to existing techniques, which generally require specially tailored model-specific QMC updates for each to-be-simulated system, and are thus limited to simulating models of very specified structures and geometries. 

We believe that the generality and versatility of our approach make our proposed technique a very useful tool for condensed matter physicists studying spin systems, allowing the community to explore, with ease, an extremely wide range of physical models, many of which have so far been inaccessible, cumbersome to code, or too large to implement with existing techniques. To that aim, we have made our program code freely accessible on GitHub~\cite{githubPMRQMC}.

We note though that while our approach guarantees a correct equilibrium distribution of the Markov chain, the proposed algorithm does not guarantee a universal rapid mixing of the Markov chain, nor does it resolve or aims to resolve the sign problem. 

The generality of the technique covered in this study makes it easily extendable to other types of systems, e.g., fermionic, bosonic or higher spin systems. We intend to explore these in future work. 

\begin{acknowledgments}
This project was supported in part by NSF award \#2210374. In addition, this material is based upon work supported
by the Defense Advanced Research Projects Agency (DARPA) under Contract No. HR001122C0063. All material, except scientific articles or papers published in scientific journals, must, in addition to any notices
or disclaimers by the Contractor, also contain the following disclaimer: Any opinions, findings and conclusions or recommendations expressed in this material are those of the author(s) and do not necessarily reflect the views of the
Defense Advanced Research Projects Agency (DARPA).
\end{acknowledgments}
\pagebreak
\bibliography{refs}
\appendix

\section{Cycle completion with gaps}
\label{appendix:cycle_completion}

As was mentioned in the main text, a simple cycle completion protocol, in which consecutive elements of $S_{{\bf{i}}_q}$
form a subsequence $S$, may lead to a violation of the ergodicity condition and therefore also an incorrect calculation.
The reason for this is the possibility of the resultant configuration having zero weight [as per Eq.~(\ref{eq:gbw})]. As a consequence, the fundamental cycle completion routine may never be applied for some of the fundamental cycles
during the Markov process, which can in turn lead to the ergodicity violation.

To resolve this issue, we have developed a protocol that we call `cycle completion with gaps'. This protocol does not require the elements of the sequence $S$ to form a consecutive unit within $S_{{\bf{i}}_q}$.
A detailed description of this subroutine follows.

The parameters are: $r_{\min}$, $r_{\max}$, $l_{\min}(r)$, $l_{\max}(r)$.
We usually choose: $r_{\min} = (f_{\min} - 1)/2$, $r_{\max} = (f_{\max} + 1)/2$, 
$l_{\min}(r)=2r-1$, $l_{\max}(r) = 2r+1$,
where $f_{\min}$ and $f_{\max}$ are minimal and maximal fundamental cycles lengths, respectively.
The other option (`exhaustive search') is to choose $r_{\min} = 0$, $r_{\max} = f_{\max}$,
$l_{\min}(r) = r$, $l_{\max}(r) = f_{\max}$.

The sequence of operations is as follows.
\begin{enumerate}
\item Pick a random integer $u$ according to a geometric distribution $p_u$. As we'll see, $u$ is the total number of operators in the `gaps'.
\item If $q < u + r_{\min}$, then the update is rejected.
\item Pick a random integer $r$ such that $r_{\min} \leq r \leq \min(r_{\max},q-u)$.
      We note that the probability $p_r(q) = (\min(r_{\max},q-u)-r_{\min}+1)^{-1}$ depends on $q$.
\item Randomly pick a sub-sequence $\widetilde{S}$ of length $r+u$ containing consecutive operators from the sequence $S_{{\bf i}_q}$.
\item Randomly choose a subsequence $S$ of length $r$ from $\widetilde{S}$. The remaining $u$ operators in $\widetilde{S}$ we will call `gaps'.
\item If $S$ contains repeated operators, the update is rejected.
\item Find all fundamental cycles of lengths $l$ such that $l_{\min}(r) \leq l \leq l_{\max}(r)$, each containing all operators 
      of the sub-sequence $S$. Denote by $n_c$ the number of found cycles.
\item If $n_c = 0$, the update is rejected. Otherwise, we randomly choose one of the found fundamental cycles.
\item Attempt to replace the sub-sequence $\widetilde{S}$ of length $r+u$ by the sequence $\widetilde{S'}$ of length $r'+u$ which contains
      all the remaining $r'$ operators from the selected cycle, as well as all the `gaps'.
      We shuffle the sequence $\widetilde{S'}$ so that its operators are contained in random order.
      We accept the update with the probability $P_{accept}$, which is considered below.
\end{enumerate}
Let us now find the acceptance probability $P_{accept}$ such that the detailed balance holds for the above protocol.
Suppose that the $u$ gaps contain $u_i$ of operators $P_i$, where $i = 1,2,\dots, M$, so that $\sum_i u_i = u$.
Let us denote the old and new configurations as $A$ and $B$, probability to select $B$ from $A$ as $P_{select}(A\to B)$,
probability to select $A$ from $B$ as $P_{select}(B\to A)$. Then, we have
\begin{multline}
P_{select}(A\to B) = p_u \cdot p_r(q) \cdot (q-(r+u)+1)^{-1} \times \\ \times \frac{1}{n_c} \cdot \binom{r+u}{u}^{-1} \binom{r'+u}{u}^{-1} \cdot \frac{1}{r'!} \cdot \frac{u_1! \dots u_M!}{u!},
\end{multline}
\begin{multline}
P_{select}(B\to A) = p_u \cdot p_r(q') \cdot (q'-(r'+u)+1)^{-1} \times \\ \times \frac{1}{n'_c} \cdot \binom{r'+u}{u}^{-1} \binom{r+u}{u}^{-1} \cdot \frac{1}{r!} \cdot \frac{u_1! \dots u_M!}{u!}.
\end{multline}
Here, $n'_c$ is the number of fundamental cycles of lengths $l$ such that $l_{\min}(r') \leq l \leq l_{\max}(r')$, each containing 
all $r'$ operators of the sub-sequence $S'$.
Since $q'=q+r'-r$, we have $q-(r+u)+1 = q'-(r'+u)+1$. Therefore,
\begin{align}
P_{accept}&(A\to B) \notag \\
&= \min\left(1,\frac{W_B}{W_A}\cdot\frac{P_{select}(B\to A)}{P_{select}(A\to B)}\right) \notag \\
&=\min\left(1,\frac{W_B}{W_A}\cdot\frac{p_{r}(q')}{p_r(q)}\cdot\frac{n_c}{n'_c}\cdot\frac{r'!}{r!}\right).
\label{eq:Paccept}
\end{align}
Here, $W_A$ and $W_B$ are the weights of the old and the new operator sequences.
Because $P(A\to B) = P_{select}(A\to B) P_{accept}(A\to B)$
and $P(B\to A) = P_{select}(B\to A) P_{accept}(B\to A)$, Eq.~(\ref{eq:Paccept}) satisfies the detailed balance condition.

\section{Estimation of the statistical errors}
\label{appendix:errors}

It is known that statistical errors in a Monte-Carlo calculation can be estimated
by employing binning analysis~\cite{efron1994book,Janke2008}.
By grouping $N$ measurements of an observable $\cal O$ into $n_B$ non-overlapping blocks of length $B = N/n_B$,
one first obtains a single data point for each of the $n_B$ bins as follows
\beq
{\cal O}_i^{(B)} = \frac{1}{B} \sum_{j=(i-1)B+1}^{iB} {\cal O}_j,\qquad i = 1,\dots,n_B.
\eeq
If the bins are large enough, the averages ${\cal O}_i^{(B)}$ will be effectively uncorrelated,
and one could then use the simple (uncorrelated) variance estimator to find the variance of the mean,
\beq
\sigma^2(\langle{\cal O}\rangle) = \frac{1}{n_B(n_B-1)} \sum_{i=1}^{n_B} ({\cal O}_i^{(B)} - \langle{\cal O}\rangle)^2,
\label{eq:variance}
\eeq
where $\langle{\cal O}\rangle = \langle{{\cal O}^{(B)}}\rangle = \sum_{j=1}^{n_B} {\cal O}^{(B)}_j / n_B$.
Similarly, the covariance estimator of the mean values of two observables $\cal O$ and $\cal Q$ is
\begin{multline}
\textrm{cov}(\langle{{\cal O}}\rangle,\langle{{\cal Q}}\rangle) = \\
\frac{1}{n_B(n_B-1)} \sum_{i=1}^{n_B} ({\cal O}_i^{(B)} - \langle{\cal O}\rangle)({\cal Q}_i^{(B)} - \langle{\cal Q}\rangle).
\label{eq:covariance}
\end{multline}
Since $\langle A\rangle$ is the ratio Eq.~(\ref{eq:Amean}) of two random variables,
the improved approximation for $\langle A\rangle$ and the approximation for $\sigma^2(A)$
are as follows~\cite{kendall1998book}:
\begin{multline}
\langle A\rangle = 
\frac{\langle A\cdot \sgn\rangle}{\langle \sgn\rangle}\left( 1 + \frac{\sigma^2(\langle\sgn\rangle)}{\langle \sgn\rangle^2}\right) -\\
\frac{\textrm{cov}(\langle\sgn\rangle,\langle A\cdot\sgn\rangle)}{\langle \sgn\rangle^2},
\end{multline}
and 
\begin{multline}
\sigma^2(A) =
\frac{\langle A\cdot \sgn\rangle^2}{\langle \sgn\rangle^2}\left(
\frac{\sigma^2(\langle A\cdot\sgn\rangle)}{\langle A\cdot \sgn\rangle^2} +\right. \\
\left.\frac{\sigma^2(\langle\sgn\rangle)}{\langle\sgn\rangle^2} -
2\frac{\textrm{cov}(\langle\sgn\rangle,\langle A\cdot\sgn\rangle)}{\langle\sgn\rangle\langle A\cdot \sgn\rangle}
\right).
\end{multline}
Here, the values of $\sigma^2(\langle\sgn\rangle)$, $\sigma^2(\langle A\cdot\sgn\rangle)$,
and $\textrm{cov}(\langle\sgn\rangle,\langle A\cdot\sgn\rangle)$ should be obtained via Eqs.~(\ref{eq:variance}) and~(\ref{eq:covariance}).

\end{document}